%% file: OIKF_journal_paper_arXiv_v3.tex
\newif\ifsingle

\newif\ifFullVersion
\FullVersiontrue 

\ifsingle
\documentclass[11pt,draftclsnofoot, onecolumn]{IEEEtran}		
\else		
\documentclass[10pt,final, twocolumn]{IEEEtran}
\fi

\usepackage{STSP}
\usepackage{NUVD}

\title{Outlier-Insensitive Kalman Filtering:\\  Theory and Applications}
\author{  
	\IEEEauthorblockN{Shunit Truzman, Guy Revach, Nir Shlezinger, Itzik Klein\\ 
	} 
	\thanks{
Parts of this work were presented at the IEEE International Conference on Acoustics, Speech, and Signal Processing (ICASSP) 2023 \cite{truzman2022outlier}. 

S. Truzman and I. Klein are with the Hatter Dept. of Marine Technologies, University of Haifa, Haifa, Israel, (e-mail: shunitruzman@gmail.com, kitzik@univ.haifa.ac.il). 
G. Revach is with the Institute for Signal and Information Processing (ISI), D-ITET, ETH Zürich, 
Switzerland (e-mail: grevach@ethz.ch). 
N. Shlezinger is with the School of ECE, Ben-Gurion University of the Negev, Be'er Sheva, Israel (e-mail: nirshl@bgu.ac.il). 

S. Truzman is supported by the Maurice Hatter Foundation. 

The authors thank Prof. Hans-Andrea Loeliger for the helpful discussions.

	}}
	
\vspace{-0.75cm}

\begin{document}
	
 \maketitle
	\pagestyle{plain}
	\thispagestyle{plain}
%
%
\begin{abstract}
State estimation of dynamical systems from noisy observations is a fundamental task in many applications. It is commonly addressed using the linear \ac{kf}, whose performance can significantly degrade in the presence of outliers in the observations, due to the sensitivity of its convex quadratic objective function. To mitigate such behavior, outlier detection algorithms can be applied. In this work, we propose a parameter-free algorithm which mitigates the harmful effect of outliers while requiring only a short iterative process of the standard \ac{kf}’s update step. To that end, we model each potential outlier as a normal process with unknown variance and apply online estimation through either expectation maximization or alternating maximization algorithms. Simulations and field experiment evaluations demonstrate our method’s competitive performance, showcasing its robustness to outliers in filtering scenarios compared to alternative algorithms.
\end{abstract}

%

\acresetall 
%
%
%
\section{Introduction}
\label{sec:Intro}
%
%
\IEEEPARstart{S}{tate} estimation from noisy observation is a core task in various signal processing applications~\cite{Durbin2012}, such as localization and tracking\cite{ZhuZLL2020, Yuan2023, NavonB21}. This task is commonly addressed by the celebrated \ac{kf}~\cite{Kalman1960}, a recursive and efficient algorithm providing an optimal low-complexity solution under the Gaussian noise and linear dynamics assumptions. However, the \ac{kf}'s performance degrades significantly when observations are impaired by outliers, due to its \acl{ls} cost function~\cite{Roncetti2009, Farahmand2011, Aravkin2013}. In \acl{rw} scenarios, measurements, especially from lower-quality sensors such as \ac{gnss} devices, often contain outliers~\cite{Wang2009,Zhu2019, NavonB21}. This presents a significant challenge to the \ac{kf} effectiveness. Therefore, an algorithm's ability to remain insensitive to outliers plays a crucial role in state estimation missions.
%
%
Various techniques were proposed in the literature to cope with outliers: basic techniques, such as those in~\cite{Ye2001, Lekkas2015, VanWyk2020}, employ statistical tests like the \acl{ch2t} to identify outliers based on prior information, and subsequently reject them. However, their robustness against outliers relies solely on the prediction step. The methods in~\cite{Ting2007, Agamennoni2011, Tao2023} suggest reweighting the observation noise covariance at each update step, but they often require extensive hyper-parameter tuning. The approaches in~\cite{karlgaard2007, gandhi2009, Farahmand2011, Aravkin2017} strive to reduce the \ac{kf}'s sensitivity to outliers by replacing its quadratic cost function. Specifically, the works~\cite{karlgaard2007,gandhi2009} propose a Huber-based \ac{kf} by minimizing the combined $\mathcal{L}_1$ and $\mathcal{L}_2$ norms. The nominal noise is bounded using a Huber function, {but the feature of heavy tails inherent in non-Gaussian noises could limit the estimation accuracy}.
The techniques in~\cite{Farahmand2011, Aravkin2017} substitute the quadratic cost function with more suitable, often nonsmooth, convex functions, controlling outliers by promoting sparsity. However, these techniques involve smoothing algorithms rather than filtering and can be computationally complex. Methods such as~\cite{Agamennoni2012} employ heavy-tailed distributions, like the \acl{sstd}, to model the observation noise. However, in the absence of outliers, a significant degradation is expected due to the violated Gaussian assumption. The authors of \cite{Huang2017} address this problem by using a hierarchical distribution, specifically adopting a more robust distribution when the noise is skewed. \textcolor{black}{Another popular technique, known as the maximum correntropy \ac{kf} (MCKF) \cite{Wang21,CHEN2017}, enhances the \ac{kf}’s performances for state estimation in presence of non-Gaussian noises, where correntropy is maximized. However, when the process model has uncertainties, the performance of MCKF degrades \cite{Liu2017}}. With recent advancements in \acp{nn}, methods such as~\cite{VanWyk2020, Davari2021} suggest detecting and correcting outlier observations using \acp{nn} before they enter into the \ac{kf} stage. Nonetheless, these methods often require access to large amounts of data and pre-training.
%
%
In~\cite{Wadehn2016}, the use of \ac{nuv} prior is introduced to devising an outlier insensitive \ac{ks}. Inspired by sparse Bayesian learning~\cite{Tipping01, 1315936}, the authors propose to model each potential outlier as \ac{nuv}~\cite{Loeliger2017, Loeliger2019, Loeliger2023}, and estimate the unknown variance using \ac{em} algorithm~\cite{EM1977,PalmerWKR05}, resulting in sparse outlier detection. 
In~\cite{Wadehn2016} focuses on the offline smoothing task and proposes only the derivation of the \ac{em} to estimate the unknown variance, which requires the computation of second-order moments. \textcolor{black}{In addition, this work focuses on a smoothing problem, commonly used for post processing, where all the data is available and one can run a forward and backward pass algorithm that allows to refine the state estimates simultaneously, which makes the task simpler compared to \ac{kf}.}

%
\textcolor{black}{In the preliminary findings of this work, reported in the conference paper \cite{truzman2022outlier}}, we introduced the \ac{oikf}, which is designed for the more commonly encountered task of online real-time filtering. 
In addition to presenting the \ac{em} algorithm, we also provide the \ac{am} algorithm~\cite{AAndresen2016} for estimating the unknown variance, which eliminates the need for computing the second-order moment of the state vector, makes its implementation simpler \textcolor{black}{and requires significantly less computation time.} \textcolor{black}{The main advantages of our approach, compared to other existing outlier-robust \ac{kf} methods, are that it  $(i)$ is parameter-free;  $(ii)$ amounts to a short iterative process within the \ac{kf}’s update step, i.e., we effectively stay within the linear Gaussian framework, and $(iii)$ effectively leverages all observation samples during the state estimation process of the \ac{kf}.} \textcolor{black}{This paper extends the preliminary findings reported in our conference paper \cite{truzman2022outlier} with the following additional contributions:
}
%
\vspace{2mm}
\begin{enumerate}
\item 	\textbf{Motivation}: A comprehensive motivation for the utilization of \ac{nuv} in outlier detection within the \ac{kf} framework, highlighting its benefits. 
\vspace{1.5mm}
\item \textbf{Theory}: {A comprehensive elucidation of the motivation behind the \ac{nuv} prior representation to model outlier and tackle the problem of state estimation in the presence of outliers. To that end, we provide a complete mathematical derivation providing in-depth insights into the theoretical aspects of our \ac{oikf} with its two implementations \ac{oikf}-\ac{em} and \ac{oikf}-\ac{am}.}
\vspace{1.5mm}
\item \textbf{Extensive Simulation Analysis}: a comparison for scenarios with low outlier intensity, suitable for any sensor updating the \ac{kf}, which are inherently more challenging to detect and compensate for. 
\vspace{1.5mm}
\item \textbf{\ac{gnss} Outlier Detection}: A real-world analysis focused on \ac{gnss} outlier detection using two datasets with three different platforms to highlight our approach's robustness. { One dataset comprises of Segway recordings~\cite{Bianco2016}, while the other comprises data from a quadrotor and a marine vessel~\cite{Shurin2022}.}
\vspace{1.5mm}
\item \textbf{Open Source}: The source code and additional information on our empirical study can be found at \textcolor{blue}{\underline{https://github.com/KalmanNet/OIKF-NUV.git}}. 
\end{enumerate}
%
%
%
The rest of this paper is organized as follows: \secref{sec:SysModel} reviews the preliminaries for the \acl{or} state estimation task. \secref{sec:OIKF} provides detailed explanations of \ac{nuv} modeling and its utilization in the \ac{oikf}. \secref{sec:NumEval} presents the results of the empirical study, while \secref{sec:Conclusions} concludes the paper with final remarks.
%
%
\section{Problem Formulation and Preliminaries}
\label{sec:SysModel}
In this section, we introduce the preliminaries for the task of \acl{or} online state estimation, namely, the \ac{ss} model with outliers, and recapitulate the \ac{kf} algorithm.
%

%
\subsection{State Space Model with Outliers}
We consider a scenario where noisy time-series observations, denoted as $\{\gmat{y}_\tau\}_{\tau =1}^{t}$, are sequentially presented to a filter. The objective is to provide a sequence of estimates, $\{\hat{\gmat{x}}_\tau\}_{\tau =1}^{t}$, corresponding to a sequence of hidden (latent) values or 'states', $\{\gmat{x}_\tau\}_{\tau =1}^{t}$~\cite{Revach2022}. This scenario introduces an additional challenge: a subset of the observations may be impaired by outliers from an unknown distribution. We operate under the assumption that an anomalous observation should be considered a rare event to qualify as an outlier. 

Unlike the offline state estimation task, \acl{aka} as smoothing, which is considered in~\cite{ni2022rtsnet}, where  all observations are provided as a batch, we focus on \acl{rt} filtering here. In this approach, the estimate of $\gmat{x}_t$ relies solely on current and past observations. This stands in contrast to the methodology in~\cite{Wadehn2016}, where iterating on the entire batch of observations is used to enhance robustness to outliers and consequently improve state estimation performance. 
 
In this work, we assume that the underlying relationship between the observed values and the hidden values is represented by a \ac{ss} model~\cite{Durbin2012}. 
We focus on a \acl{lg} \ac{ss} model in \acl{dt}, $t\in\mathbb{Z}$, represented as follows:
\begin{subequations}
\begin{align}
    \label{eq:ssmodelx}
    \gvec{x}_{t}&= 
    \gvec{F}\cdot{\gvec{x}_{t-1}}+\gvec{e}_{t},
    \quad
    \,\,\,\,\,\,\gvec{e}_t\sim 
    \mathcal{N}\brackets{\gvec{0},\gvec{Q}},
    \quad
    \gvec{x}_{t}\in\greal^m\\
    \label{eq:ssmodely}
    \gvec{y}_{t}&=
    \gvec{H}\cdot\gvec{x}_t+\gvec{z}_t+\gvec{u}_t,
    \,\,\,\,\gvec{z}_t\sim
    \mathcal{N}\brackets{\gvec{0},\gvec{R}},
    \quad
    \gvec{y}_{t}\in\greal^n.
\end{align}
\label{eq:ssmodel}
\end{subequations}
\geqref{eq:ssmodelx} describes the time evolution of the state $\gvec{x}_{t}$ from the previous state $\gvec{x}_{t-1}$, governed by an system (evolution) matrix $\gvec{F}$ and \acl{agn} $\gvec{e}_{t}$. This noise, with a process covariance matrix $\gvec{Q}$, represents potential modeling uncertainties. \geqref{eq:ssmodely} portrays how observations $\gvec{y}_{t}$ are generated from $\gvec{x}_{t}$, the current state at time step $t$. This process involves a measurement  (observation) matrix $\gvec{H}$, \acl{agn} $\gvec{z}_{t}$, with a measurement covariance matrix $\gvec{R}$ accounting for uncertainties in the measurements, and potential outliers $\gvec{u}_t$, which follows an unspecified distribution. 
%
%

\subsection{Linear Kalman Filtering}
The celebrated \ac{kf}~\cite{Kalman1960} is particularly noteworthy for its recursive and efficient algorithm, providing an optimal solution under Gaussian noise and linear dynamics~\cite{Bar2004, Durbin2012}. In its most general form, the \ac{kf} aims to estimate the current state based on a noisy observation signal. However, the \ac{kf}'s performance can degrade in the presence of outliers~\cite{Farahmand2011, Aravkin2013, Aravkin2017}. This sensitivity stems from the filter's objective to minimize a quadratic cost function, a structure that inherently is not able to follow fast jumps in the state dynamics\cite{OhlssonGLB12}. For full details on how the \ac{map} formulation boils down to \acl{ls} minimization, see\cite{Bell1993, Humpherys2012}.

The \ac{kf} estimates the state $\gvec{x}_t$ from the observations $\set{\gvec{y}_\tau}_{{\tau\leq t}}$ and 
can be thought of as a two-step process at
each time step: \emph{predict} and \emph{update}. In the  \emph{predict} step, the joint probability distribution is computed using the first and second-order moments of the Gaussian distribution, resulting in the
prior distribution.
%
%
The \emph{predict} of the $1^{\textrm{st}}$ and $2^{\textrm{nd}}$ order moments:
\begin{subequations}
\begin{align}
\label{eqn:predict}
    \hat{\gvec{x}}_{t\given{t-1}} &= 
    \gvec{F}\cdot{\hat{\gvec{x}}_{t-1}},
    \hspace{0.45cm}
    \mySigma_{t\given{t-1}} \!=\!
    \gvec{F}\cdot\mySigma_{t-1}\cdot\gvec{F}^\top\!+\!\gvec{Q},\\
    \label{eqn:predict2}
    \hat{\gvec{y}}_{t\given{t-1}} &=
    \gvec{H}\cdot{\hat{\gvec{x}}_{t\given{t-1}}},
    \hspace{0.2cm}
    \gvec{S}_{t\given{t-1}}\! =\!
    \gvec{H}\cdot\mySigma_{t\given{t-1}}\cdot\gvec{H}^\top\!+\!\gvec{R}.
\end{align}
\end{subequations}
where $\mySigma$ represents the covariance of the state, $\gvec{F}$ is the state-transition model, and $\gvec{H}$ is the observation and model. The matrices $\gvec{Q}$ and $\gvec{R}$ are the covariance matrices of the process noise and observation noise, respectively.
The \ac{kf} uses this prior distribution in the \emph{update} step in the \emph{posterior}  distribution calculation by computing the new observation $\gvec{y}_t$ with the previously predicted \emph{prior} $\hat{\gvec{x}}_{t\given{t-1}}$.
%
%
And the \emph{update} of the $1^{\textrm{st}}$ and $2^{\textrm{st}}$ order statistical moment
\begin{equation}
    \hat{\gvec{x}}_{t} = 
    \hat{\gvec{x}}_{t\given{t-1}}+\Kgain_{t}\cdot\Delta\gvec{y}_t,\hspace{-0.24cm}
    \quad
    {\mySigma}_{t}\! =\!
    {\mySigma}_{t\given{t-1}}\!-\!\Kgain_{t}\cdot{\mathbf{S}}_{t\given{t-1}}\cdot\Kgain^{\top}_{t},
\label{eqn:update}
\end{equation} 
%
%
\begin{equation}\label{eq:kgain}
    \Kgain_{t}={\mySigma}_{t\given{t-1}}\cdot{\gvec{H}^\top}\cdot{\gvec{S}}^{-1}_{t\given{t-1}}, 
    \quad
    \Delta\gvec{y}_t=\gvec{y}_t-\hat{\gvec{y}}_{t\given{t-1}}.
\end{equation}
where $\Kgain_t$ is the Kalman gain matrix used to balance the contributions of both parts and produce the final posterior distribution.
%
%
\section{Outlier-Insensitive Kalman Filtering Using NUV Prior} \label{sec:OIKF}
{This section introduces our \ac{oikf} algorithm. \textcolor{black}{First, we present the particular property of the \ac{nuv} prior that motivates us to model outliers as \ac{nuv} and helps us tackle the problem of state estimation in the presence of outliers, as seen in \ac{kf}.} Then, we elaborate on our innovative approach of integrating the \ac{nuv} prior into the \ac{kf} algorithm for outlier detection, denoted as \ac{oikf}. Finally, we provide a comprehensive derivation of our two proposed algorithms to estimate the unknown variance of the \ac{nuv}, namely  \ac{nuv}-based \ac{em} and \ac{nuv}-based \ac{am}.}
%
%
\subsection{Motivation For NUV Prior 
Representation}\label{subsec:NUV_modeling}

The \ac{nuv} formulation models a variable of interest as a normal distribution with unknown variance, given that the unknown variance has a prior distribution~\cite{Loeliger2017,Loeliger2023}. {The \ac{nuv} representation method proves to be a robust approach with various applications, each encountering different problems, and the choice of a specific prior depends on its circumstances. For example, \cite{Ma2017} proposes different priors for computer imaging problems. \textcolor{black}{In our specific problem, outlier-insensitive \ac{kf}, we opt for a uniform prior, based on a previous work \cite{Loeliger2019}, which demonstrates scenarios similar to our \ac{ss} model.} This choice is motivated by computational convenience and the objective of resulting in sparse outlier detection, as explained in \ssecref{subsec:EM_OIKF} and in \ssecref{subsec:discussion}. Once the prior is set, our approach is in fact a parameter free approach.}

One well-known property of the \ac{nuv} is its tendency to yield a non-convex penalty~\cite{Loeliger2019},  suitable for addressing sparse least-squares problems with outliers.
This non-convex penalty is motivated by the influence function~\cite{RousseeuwH2011} in residuals. This function assesses the effect of a residual's size on the loss by evaluating its derivative~\cite{Aravkin2017}. As the size of the residuals increases, the influence function gradually approaches zero, leading to a sparse solution.

%
To motivate, we employ a simple example, based on the observation model~\eqref{eq:ssmodely}, which illustrates the fundamental property of \ac{nuv} priors. Consider a single observation of the form
\begin{equation}
    \gscal{y}=\gscal{v}+\gscal{u},\quad \gscal{y}\in\greal
\end{equation}
Here, $\gscal{v}$ is an \ac{awgn} with variance $\gscal{r}^2$. The variable of interest $\gscal{u}$ is modeled as a zero-mean real scalar Gaussian random variable with an unknown variance $\gamma^2$ (\ac{nuv}).
The \ac{mle} of $\gamma^2$ from a single sample $\gscal{y} \in \mathbb{R}$ can be computed as follows:
{
\begin{equation}
\hat{\gamma}^2 = \argmax_{\gamma^2\geq0}\prob{\gscal{y}\given{\gamma^2}}\prob{\gamma^2}\label{eqn:gamma_MLE,10}
\end{equation}
when we assume a constant prior $\prob{\gamma^2}=\sqrt{2\pi}$ for computation convenience and $\prob{\gscal{y}\given{\gamma^2}}$ is normally distributed, that is $\prob{\gscal{y}\given{\gamma^2}}\sim\mathcal{N}\brackets{0,\gscal{r^2+\gamma^2}}$, then \eqref{eqn:gamma_MLE,10} can be rewritten as: 
\begin{equation}
\hat{\gamma}^2=\argmax_{\gamma^2\geq0}\set{\frac{1}{\sqrt{2\pi\brackets{\gscal{r}^2+\gamma^2}}}  \exp \brackets{\frac{-\gscal{y}^2}{2(\gscal{r}^2+\gamma^2)}}}\label{eqn:gamma_MLE,11}
\end{equation}
To simplify the computation of \eqref{eqn:gamma_MLE,11} is written in terms of a logarithmic function:}

{\begin{equation}
  \hat{\gamma}^2 = \argmin_{{\gamma^2}\geq0}  \set{\ln \brackets{\gscal{r}^2+\gamma^2} + \frac{{\gscal{y}^2}}{\brackets{ \gscal{r}^2+\gamma^2}}}.
    \label{eqn:gamma_MLE,12}
\end{equation}}
In order to derive \eqref{eqn:gamma_MLE,12}, we equate its derivative with regard to ${\gamma^2}$ to  zero and we get the closed form of unknown variance $\gamma^2$ of Gaussian $\gscal{u}$:
\begin{equation}
    \hat{\gamma}^2= \max \set{\gscal{y}^2-\gscal{r}^2,0}.
 \label{eqn:gamma_MLE,2}
\end{equation}
In a subsequent step, assuming $\gamma^2$ is estimated as $\hat{\gamma}^2$ as in~\eqref{eqn:gamma_MLE,2}, the \ac{map} estimate of $\gscal{u}$, denoted as $\hat{\gscal{u}}$, is given by:
{\begin{equation}
\hat{\gscal{u}}=\argmax_{\gscal{u}}\prob{\gscal{y}\given{\gscal{u}}}\cdot\prob{\gscal{u}} 
\end{equation}
when $\prob{\gscal{y}\given{\gscal{u}}}\sim\mathcal{N}\brackets{\gscal{u},\gscal{r}^2}$ and \textcolor{black}{$\prob{\gscal{u}}\sim\mathcal{N}\brackets{0,\gamma^2}$}, the derivation of $\hat{\gscal{u}}$ can be accomplished simper to $\gamma^2$ in \eqref{eqn:gamma_MLE,2}, thus,
 \begin{equation}
\begin{aligned}      \hat{\gscal{u}}=&\argmax_{\gscal{u}}\set{\frac{1}{\sqrt{2\pi\gscal{r}^2}}e^\frac{-\brackets{\gscal{y}-\gscal{u}}^2}{2\gscal{r}^2}+\frac{1}{\sqrt{2\pi\gamma^2}}e^{\frac{-\gscal{u}^2}{2\gamma^2}}}\\
      =&\underset{\gscal{u}}{\arg\min}\, \set{\frac{\brackets{\gscal{y}-\gscal{u}}^2}{2\gscal{r}^2}+\frac{{\gscal{u}}^2}{2\gamma^2}}.
\label{eqn:u_MLE,1} 
\end{aligned}    
\end{equation}}
Maximizing the expression in~\eqref{eqn:u_MLE,1}
results in
\begin{align} 
    \gscal{\hat{u}}=&\gscal{y}\cdot\frac{\hat{\gamma}^2}{\hat{\gamma}^2+\gscal{r}^2} =  \max \set{ \frac{\gscal{y}^2-\gscal{r}^2}{\gscal{y}},0}.
\label{eqn:u_MLE,2} 
\end{align}
%
%
Plugging the obtained result  \eqref{eqn:gamma_MLE,2} into $\prob{\gscal{y}}$:
\begin{align}
    \prob{\gscal{y}}=\max_{\gamma^2\geq0} \prob{\gscal{y}\given\gamma^2}\cdot \prob{\gamma^2}
    \label{eqn:p(y)}
\end{align}
{yields the equivalent cost function $\loss{\gscal{y}}=-\log\,\prob{\gscal{y}}$, when $\prob{\gscal{y}}$ can be derived from \eqref{eqn:gamma_MLE,12} when we also assume a constant prior for $\prob{\gamma^2}$, and we obtain:
\textcolor{black}{
\begin{equation}
\begin{aligned}
    \loss{\gscal{y}} =&
    -\log\sbrackets{\frac{1}{\sqrt{\gscal{r}^2+\gamma^2}}\exp\brackets{\frac{-y^2}{2\brackets{\gscal{r^2}+\gamma^2}}}}\\
    =&\frac{1}{2}\log\brackets{\gscal{r^2}+\gamma^2}+\frac{\gscal{y^2}}{2\brackets{\gscal{r^2+\gamma^2}}}
\label{eqn:lossfunc1}
\end{aligned}
\end{equation}}
Using the obtained expression for $\hat{\gamma}^2$ from \eqref{eqn:gamma_MLE,2}, we obtain:
\begin{equation}
\begin{aligned}
    \loss{\gscal{y}} =
    \begin{cases}
    \gscal{y}^2/\brackets{2\gscal{r}^2}+\log\,\gscal{r}, & {\gscal{y}^2}< \gscal{r}^2\\
    \log\abs{\gscal{y}}+0.5, &{\gscal{y}^2}\geq \gscal{r}^2
    \end{cases}
\label{eqn:lossfunc}
\end{aligned} 
\end{equation}
If $\gscal{r}>0$, \eqref{eqn:lossfunc} results in a nonconvex function, which proves valuable for handling sparse \acl{ls} models, such as \ac{kf}, in the presence of outliers\cite{Loeliger2019}.} {Through this example, we establish that $\gamma^2=0$ leads to $\gscal{u}=0$, indicating that no outlier is identified and the obtained $\gamma^2$ leads to a nonconvex cost function.}

%
\subsection{Kalman Filtering with NUV Prior}\label{subsec:KF_NUV}
\begin{figure}[t]
    \centering
    \includegraphics[scale=0.65]{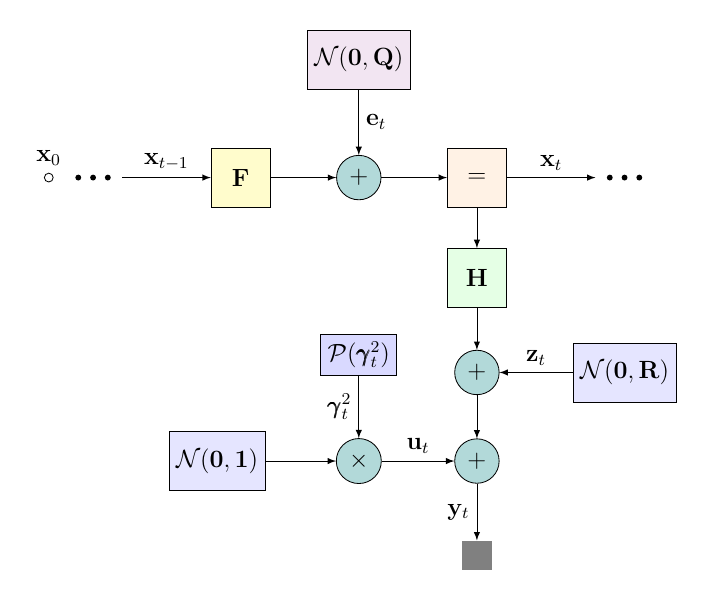}
    \caption{Factor graph of the system model at time step $t$}\label{fig:SSmodel}
    \vspace{3mm}
\end{figure}

The proposed \ac{oikf} introduces a new form of the \ac{ss} model~\eqref{eq:ssmodel} by incorporating an additional variable into the observation signal, denoted as $\gvec{u}_t$. This variable $\gvec{u}_t$ represents the impulsive noise responsible for causing outliers. To improve the model's capability to handle heavy-tailed distributions in observations, we model this outlier $\gvec{u}_t$ as \acl{nuv} ${\vgamma}^2_t$, namely:
\begin{equation}
    \gvec{u}_t\sim \mathcal{N}(0,{\vgamma}^2_t), \quad {\vgamma}^2_t\sim \prob{{\vgamma}^2_t}
    \label{eqn:nuv}
\end{equation}
where $p({\vgamma}^2_t)$ is the prior distribution of ${\vgamma}^2_t$. \textcolor{black}{The assumption that \ac{nuv} modeling follows a Gaussian distribution  does not imply that the noise obeys a conventional \ac{iid} Gaussian model and that it is perfectly suitable for representing outliers. This approach allows us to remain within the linear Gaussian framework without making strict assumptions about the outlier distribution.}
\figref{fig:SSmodel} demonstrates visually the integration of the \ac{nuv} representation into the overall model through a factor graph. 
The \ac{nuv} representation approach results in a sparse outlier detection solution~\cite{Loeliger2017,Loeliger2023}, indicating that most values of $\vgamma^2_t$ will be zero. Consequently, as can be seen in \eqref{eqn:u_MLE,2} this leads to $\gvec{u}_t=0$, 
suggesting the absence of outliers, as expected. \\
%
%
For any given observation sample $\gvec{y}_{t}$ \eqref{eq:ssmodely}, we define $\gvec{v}_t$ to be the error vector as the sum of two independent sources: the observation noise $\gvec{z}_t$ and the outlier noise $\gvec{u}_t$. 
{Thus, the covariance matrix of $\gvec{v}_t$ is equal by definition:
\begin{equation}
    \label{eqn:v_error} \gvec{v}_t\triangleq\gvec{y}_t-\gvec{H}\cdot\gvec{x}_t=\gvec{z}_t+\gvec{u}_t,
    \quad
\gvec{v}_t\sim\mathcal{N}\brackets{\gvec{0},\gvec{\Gamma}_t}
\end{equation}
$\gvec{\Gamma}_t$ is diagonal and comprises the sum of variances of the two noise sources, namely:
\begin{equation}
    \gvec{\Gamma}_t=
    \textrm{diag}\brackets{{\vnu}^2_t},
    \quad
    {\vnu}^2_t\triangleq\gvec{r}^2+{\vgamma}^2_t.
    \label{eqn:Gamma_cov}
\end{equation}}
\textcolor{black}{In case the matrix $\gvec{\Gamma}_t$ is non-diagonal, the \ac{kf} becomes nonlinear, requiring the use of an extended \ac{kf} (EKF).}
In each \ac{kf} iteration, the temporary estimate of $\vgamma^2_t$ is incorporated into the overall covariance  $\gvec{\Gamma}_t$, effectively reweighting the covariance noise of the observations. Consequently, this affects the Kalman gain $\Kgain_t$ in the \textit{update} equations, allowing to extract the information from all the noisy observations and leverage their information effectively.

For the process of \ac{map} estimation of the unknown variance,  ${\vgamma}^2_t$, we can apply either \ac{em} (\ssecref{subsec:EM_OIKF}) or \ac{am} (\ssecref{subsec:AM_OIKF}) algorithms. While in the \ac{em} approach, the second-order moment $\vnu_t$~\eqref{eqn:Gamma_cov} is directly estimated, in the \ac{am} approach, it is obtained from estimating the first-order moment $\gvec{v}_t$~\eqref{eqn:v_error} only, as summarized below: 
\begin{equation}\label{eq:nuv_est}
    {\hat{\vgamma}}^2_t=\max\set{\vnu_t^2-\gvec{r}^2,0},
    \hspace{0.175cm}
    \vnu_t^2=\set{\textrm{EM}:\hat{\vnu}_t^2,\textrm{AM}:\gvec{\hat{v}}^2_t}.
\end{equation}
For both approaches, an outlier is detected when $\hat{\vgamma}^2_t\neq0$, otherwise,  when $\hat{\vgamma}^2_t=0$, it implies no outlier is present and we revert to the standard \ac{kf}, preserving its optimally for data without outliers.
%
%
%
\subsection{Expectation Maximization}\label{subsec:EM_OIKF}
For an observation $\gvec{y}_t$, and the state vector $\gvec{x}_t$ as defined in \eqref{eq:ssmodel},  the \ac{map} estimation for the unknown variance $\hat{\vgamma}^2_t$ is
\begin{align} 
\hat\vgamma^2_{t}\brackets{\gvec{y}_t}&=\underset{\vgamma^2_{t}\geq0}{\arg\max}\,\prob{\vgamma^2_t\given{\gvec{y_t}}}=\underset{\vgamma^2_{t}\geq0}{\arg\max}\,\prob{\gvec{y}_t\given{{\vgamma^2_{t}}}}\cdot \prob{{\vgamma^2_{t}}}\\
     &=\underset{\vgamma^2_{t}\geq0}{\arg\max}\,\int\prob{\gvec{y}_t,\gvec{x}_t\given{{\vgamma^2_{t}}}}dx\cdot \prob{{\vgamma^2_{t}}}
\label{eqn:gammaEM} 
\end{align}
To solve the optimization problem in~\eqref{eqn:gammaEM}, we devise the iterative \ac{em} algorithm, which consists of two iterating steps, namely, \emph{E-step} and \emph{M-step}. \\
The \emph{E-step} determines the conditional expectation:
\begin{equation}
\begin{aligned}
    &\expectation_{\gvec{x}_t\given{\brackets{{\vgamma^{i-1}_t}}^2,\gvec{y}_t}}\sbrackets{\log\brackets{\prob{{\gvec{y}_t,\gvec{x}_t}\given \vgamma_t^2}\cdot\prob{\vgamma^2_t}}}=\\
    & \expectation_{\gvec{x}_t\given{\brackets{{\vgamma^{i-1}_t}}^2,\gvec{y}_t}}\sbrackets{\log{\prob{\gscal{y}_t\given{\gvec{x}_t,\vgamma^2_t}}}+\prob{\gvec{x}_t\given{\vgamma^2_t}}+\log{\prob{\vgamma^2_t}}}
\end{aligned}
\label{eqn:E-step}
\end{equation}
Using the \textit{Markov} property and the structure of the \ac{ss} model, the term $\prob{\gvec{x}_t\given{\vgamma^2_t}}$ is can be rewritten as $\prob{\gvec{x}_t\given{\gvec{x}_{t-1}}}$.

The \emph{M-step} goal is to maximize~\eqref{eqn:E-step} with respect to ${\vgamma^2_t}$.
In this problem, we assume a uniform prior on the unknown variance~\cite{Loeliger2017}:
\begin{equation}
    \prob{{\vgamma^2_t}}\propto 1
    \label{plain NUV}
\end{equation}
The choice of $\prob{{\vgamma^2_t}}$ to be uniform is one of many options.
As stated in~\cite{Loeliger2019}, a uniform prior on $\prob{{\vgamma^2_t}}$, also known as \emph{plain \ac{nuv}}, eventually leads to a non-convex cost function which results in the sparse effect of the unknown variance, with most of them being zeros, as expected.

Since $\prob{{\vgamma^2_t}}$ and the evolution $\prob{\gvec{x}_t\given{\gvec{x}_{t-1}}}$ do not depend on ${\vgamma^2_t}$, they can thus be omitted from the optimization process \eqref{eqn:E-step} and we can evaluate the standard \ac{em} and compute the conditional distribution in~\eqref{eqn:E-step}. Thus, the $i$th iteration step is derived by:
\begin{align} 
    \notag&Q\sbrackets{\brackets{\vgamma^{i}_t}^2}= \expectation_{\gvec{x}_t\given{\brackets{\vgamma^{i-1}_t}^2,\gvec{y}_t}}\sbrackets{\log{\prob{\gvec{y}_t\given{\gvec{x}_t,\vgamma^2_t}}}}\\ 
    \notag
    =& \expectation_{\gvec{x}_t\given{\brackets{\vgamma^{i-1}_t}^2,\gvec{y}_t}}\sbrackets{\log\brackets{\frac{1}{\sqrt{\gvec{\Gamma}_t}}\cdot\mathrm{exp}\brackets{\frac{-\brackets{\gvec{y}_t-\gvec{H}\cdot\gvec{x}_t}^2}{2\cdot \gvec{\Gamma}_t}}}}\\
     \propto& \log{\gvec{\Gamma}_t}+\frac{1}{{\gvec{\Gamma}_t}}\cdot\expectation_{\gvec{x}_t\given{\brackets{\vgamma^{i-1}_t}^2},\gvec{y}_t}\sbrackets{\brackets{\gvec{y}_t-\gvec{H}\cdot\gvec{x}_t}^2}.
    \label{eqn:EM} 
\end{align}
To expand the term $\expectation_{\gvec{x}_t\given{\brackets{\vgamma^{i-1}_t}^2},\gvec{y}_t}\sbrackets{\brackets{\gvec{y}_t-\gvec{H}\cdot\gvec{x}_t}^2}$,we utilize the \ac{kf}, allowing to obtain the first- and second-order \emph{posterior} moments of $\gvec{x}_t$, {which equal by definition:}
\begin{subequations}
\begin{align}
    &\expectation_{\gvec{x}_t\given{\brackets{\vgamma^{i-1}_t}^2,\gvec{y}_t}}\brackets{\gvec{x}_t}\triangleq\hat{\gvec{x}}_{t}^i, \label{eqn:momentIIa}\\
    &\expectation_{\gvec{x}_t\given{\brackets{\vgamma^{i-1}_t}^2,\gvec{y}_t}}\brackets{\gvec{x}_t\cdot\gvec{x}_t^\top}=\mySigma^i_{t}+\gvec{X}_t^\text{II}\triangleq\EmMoment.
    \label{eqn:momentIIb}
\end{align}
\end{subequations}
\textcolor{black}{The \ac{kf} defines $\hat{\gvec{x}}_t$ as the \emph{posteriori} state estimate mean of ${\gvec{x}}$  at time $t$, considering all observations up to and including time $t$ ~\cite{Kalman1960}. Therefore, given ${\gvec{y}_t}$ and $\brackets{\vgamma^{i-1}_t}^2$, \eqref{eqn:momentIIa} holds true. Equation \eqref{eqn:momentIIb} represents the squared first moment of ${\gvec{x}_t}$, as defined in \eqref{eqn:symbols}, plus $\hat{\mySigma}_{t}$, the \emph{posteriori} estimated covariance matrix of ${\gvec{x}_t}$, derived from the \ac{kf}~\cite{Kalman1960}.}

To simplify computations, {the following expressions are equal by definition:}
\begin{equation}
\begin{aligned}
    &\gvec{X}_t^\text{II}\triangleq\gvec{\hat{x}}^i_t\cdot {\gvec{\hat{x}}^i_t}^\top, \quad \gvec{Y}_t^\text{II}\triangleq\gvec{y}_t\cdot \gvec{y}_t^\top, \\ &\gvec{XY}_t\triangleq\gvec{\hat{x}}^i_t\cdot \gvec{y}_t^\top,  \quad \gvec{YX}_t\triangleq\gvec{y}_t\cdot{\gvec{\hat{x}}^i_t}^\top,
    \label{eqn:symbols}
\end{aligned}   
\end{equation}
Finally, the expectation step~\eqref{eqn:EM} is reduced to the following expression:
\begin{equation}    Q\sbrackets{\brackets{\vgamma^{i}_t}^2}=\log{\gvec{\Gamma}_t}+\frac{1}{{\gvec{\Gamma}_t}}\cdot\gvec{V}_t\label{eqn:EM,4}
\end{equation}
where {$\gvec{V}_t$ is equal by definition}:
\begin{equation}
    \gvec{V}_t \triangleq \set{
   \gvec{Y}_t^\text{II}-\gvec{H}\cdot\gvec{XY}_t-\gvec{YX}_t\cdot\gvec{H}^\top-\gvec{H}\cdot\EmMoment\cdot\gvec{H}^\top}
   \label{eqn:V}
\end{equation}
In \textit{M-Step}, we maximize~\eqref{eqn:EM,4} w.r.t. to $\gvec{\Gamma}_t$, thus for the $i$-th iteration:
\begin{equation}
    {\hat{\gvec{\Gamma}}^{i}_t}=\argmax_{\gvec{\Gamma}^2_t\geq0}\biggl\{\ln{\gvec{\Gamma}_t}+
    \frac{1}{\gvec{\Gamma}_t} \cdot \gvec{V}_t \biggl\}=\gvec{V}_t
   \label{eqn:Gamma^2}
\end{equation}
We further exploit the fact that $\gvec{{\Gamma}}_{t}$ in \eqref{eqn:Gamma_cov} is diagonal, to expand $\gvec{{\Gamma}}_{t}$ to its components and estimate the variance $\brackets{\gscal{\hat{\nu}}^{i}_{t,k}}^2$ for each dimension $ k\in\{1,\dots,n\}$ in a scalar manner using~\eqref{eqn:V} 
\begin{equation}
    \brackets{\hat{\nu}^{i}_{t,k}}^2=\gvec{y}_{t,k}^2-2\cdot{\gvec{y}_{t,k}}\cdot\gvec{H}\cdot\hat{\gvec{x}}_{t,k}^{i}+\gvec{H}\cdot\brackets{\hat{\gvec{x}}_{t,k}^{i}}^{\textrm{II}}\cdot\gvec{H}^\top
    \label{eqn:nu}
\end{equation}
when $\hat{\gscal{x}}_t$ is the \textit{posteriori} state estimate. From~\eqref{eqn:Gamma_cov}, \eqref{eqn:nu} and the fact that variance must be positive, we can calculate ${\gamma}_{t,k}^2$ in the $i$th iteration:
\begin{equation}
\brackets{{\hat\gamma}^{i}_{t,k}}^2=\max\set{\brackets{{\hat\nu}^{i}_{t,k}}^2-\gscal{r}_k^2,0}.
\label{eq:gammaEM}
\end{equation} 
Thus, when an outlier is detected at a time step $t$, $\gamma^2_{t,k} > 0$ otherwise its $0$, which may lead to a sparse solution.
As a consequence, the outlier will be estimated as $\gscal{\hat u}_{t,k}=0$. 

The above procedure is repeated iteratively for a fixed $\noiterations$ iterations, or until convergence is achieved. Algorithm~\ref{alg:OIKF-EM} provides the suggested \textit{pseudo-code} for the \ac{oikf} based \ac{nuv}-\ac{em}. 
%
%
\setlength{\textfloatsep}{3pt}
\begin{algorithm}
\begin{small}
\caption{\ac{oikf} based \ac{nuv}-\ac{em} for time instance $t$}\label{alg:OIKF-EM}
    \begin{algorithmic}[1]{
        \State Number of iterations $\noiterations$
        \State \textbf{Predict:} Estimate \emph{a priori} for $\gvec{\hat x}_{t\given{t-1}
        }^{i=0}, \gvec{ \mySigma}_{t\given{t-1}}^{i=0}$ via~\eqref{eqn:predict}
        \For{$i = 0,...,\noiterations-1$}
        \State
        \textbf{EM:} Estimate $\brackets{\hat\vgamma^{i}_{t}}^2$ via~\eqref{eq:gammaEM} with the $2$nd-order  \\\hspace{1.3cm} moment $\EmMoment$ as in~\eqref{eqn:momentIIb}
        \State Compute $\gvec{\Gamma}_{t}^i=\textnormal{diag}\brackets{\gvec{r}^2+\brackets{\hat\vgamma^i_{t}}^2}$ 
        \State Compute $ \gvec{\hat y}_{t\given{t-1}}^i,\gvec{S}_{t\given{t-1}}^i$ via ~\eqref{eqn:predict2}  with $\gvec{R}=\gvec{\Gamma}_{t}^i$.
        \State \textbf{Update:} Estimate \emph{a posteriori} for $\gvec{\hat x}_{t}^i, \mySigma_{t}^i$ via~\eqref{eqn:update} 
        \EndFor}
    \end{algorithmic}
\end{small}
\end{algorithm}
%
%
\subsection{Alternating Maximization}\label{subsec:AM_OIKF}
As in the EM approach, our goal is to estimate $\vgamma^2_t$ but here using \ac{am} instead. To that end, we employ the iterative \ac{am} algorithm based on the plain smoothed \ac{nuv} \cite{Loeliger2019} to compute the joint \ac{map} estimate for $\vgamma^2_t$, when the variable of interest is $\gvec{v}_t$. 
Consider the use of a \ac{nuv} prior on variable $\gvec{v}_t$ in a \ac{ss} model with observation $\gvec{y}_t$, we aim to determine their joint MAP estimate:
\begin{align}
\notag
[\hat{\gvec{v}}_{t},\hat{{\vgamma}}^2_{t}](\gvec{y}_{t})
        &=\argmax_{\gvec{v}_t,\vgamma^2_t\geq0}
        \prob{\gvec{y}_t, \gvec{v}_t, \vgamma^2_t}\\
&\!=\!\underset{{\gvec{v}_{t},\vgamma^2_{t}\geq0}}{\arg\max}\,\prob{{\gvec{y}_{t}\given{\gvec{v}_{t}}}}\cdot \prob{\gvec{v}_{t}\big|{\vgamma^2_{t}}}\cdot \prob{\vgamma^2_{t}}.
    \label{eqn:AMMap} 
\end{align}
The latter is valid because as for certain continuous random variables, the joint probability density function is defined as the derivative of the joint cumulative distribution function.

To compute \eqref{eqn:AMMap}, we derive the \ac{am} algorithm, which iterates between a maximization step over the error state ${\gvec{v}}_{t}$ with a fixed variance ${\vgamma^2_{t}}$:
\begin{equation}
    \hat{\gvec{v}}_{t}\!=\!\underset{{\gvec{v}_{t}}}{\arg\max}\, \prob{{\gvec{y}_{t}\given{\gvec{v}_{t}}}}\cdot \prob{\gvec{v}_{t}\given\vgamma^2_{t}}
    \label{eqn:VMap}
\end{equation}
In particular, we replace $\gvec{v}_{t}$ in \eqref{eqn:AMMap} with its instantaneous estimate $\gvec{\hat v}_{t}^i=\gvec{y}_t-\gvec{H}\cdot\gvec{\hat x}_{t}^i$, which can be extracted from the \ac{kf}.
The next step in the \ac{am} is maximization \eqref{eqn:AMMap} over the unknown variance ${\vgamma^2_{t}}$ based on $\gvec{v}_{t}$, resulting in
\begin{equation}
    \hat{\vgamma}^2_{t}=
    \underset{{\vgamma^2_{t}}\geq 0}{\arg\max}\,\prob{\gvec{v}_{t}\given \vgamma^2_{t}}\cdot\prob{\vgamma^2} .
    \label{eqn:GammaMap}
\end{equation}
Note that since $\prob{{\gvec{y}_{t}\given{\gvec{v}_{t}}}}$ doesn’t depend on ${\vgamma}^2_{t}$, and as we assume a uniform prior for $\prob{\vgamma^2_{t}}$ as in~\eqref{plain NUV},  these expressions are not relevant for the maximization process~\eqref{eqn:GammaMap} and can be omitted.

For convenience, we formulate~\eqref{eqn:GammaMap} in a scalar manner, which extends to multivariate observations. To do that, we assume that the observation noise $\gvec{z}_t$ and the outlier $\gvec{u}_t$ in each
dimension $k$ are independent, allowing  to treat their sum, $\gvec{v}_{t,k}$, in dimension $k$ as a scalar, leading to the scalar rule:
\begin{equation}
 {\hat{\gamma}^2_{t,k}}=\argmax_{{\gamma_{t,k}^2}\geq0}  \prob{\gscal{v_{t,k}}  |{\gamma_{t,k}^2}}
    \label{gamma_map_AM,2}
\end{equation}
This maximization rule for $\gamma^2$ is to the one in~\eqref{eqn:gamma_MLE,10}, when here the variable of interest is the Gaussian $\gscal{v}_{t,k}$. Therefore, we can use the result obtained in~\eqref{eqn:gamma_MLE,2}, and find the closed-form expression for the unknown variance of the $i$th iteration and for the $k$th entry:
\begin{equation}
    \brackets{\hat{\gamma}^{i}_{t,k}}^2 = \max \set{\brackets{\gscal{\hat{v}}^i_{t,k}}^2-\gscal{r}_{k}^2,0}
    \label{eqn:gammaAM}
\end{equation}
We obtain an analytic expression of $\hat\gamma^2_{t,k}$ for the \textit{update} step, where we alternate between $\gvec{\hat{v}}_{t}^i$ and $\hat\gamma^2_{t,k}$ until convergence. Equation \eqref{eqn:gammaAM} is parameter-free and solely relies on the \textit{posterior} estimate of the state $\hat{\gvec{x}}_{t}^i$. 
This is in contrast to \ac{em}, which incorporates estimates of both the first- and second-order moments of the states, represented as $\hat{\gvec{x}}_{t}^i$ and $\EmMoment$, \eqref{eqn:momentIIa} and \eqref{eqn:momentIIb}, respectively.
Similar to \ac{em}, this procedure is repeated iteratively. Algorithm~\ref{alg:OIKF-AM} provides the suggested \textit{pseudo-code} for the \ac{oikf} based \ac{nuv}-\ac{am}. 

%
\setlength{\textfloatsep}{3pt}
\begin{algorithm}[H]
\begin{small}
\caption{\ac{oikf} based \ac{nuv}-\ac{am} for time instance $t$}\label{alg:OIKF-AM}
    \begin{algorithmic}[1]{
        \State Number of iterations $\noiterations$
        \State \textbf{Predict:} Estimate \emph{a priori} for $\gvec{\hat x}_{t\given{t-1}
        }^{i=0}, \gvec{ \mySigma}_{t\given{t-1}}^{i=0}$ via~\eqref{eqn:predict}
        \For{$i = 0,...,\noiterations-1$}
         \State
        \textbf{AM:} Compute $\gvec{\hat v}_{t}^i=\gvec{y}_t-\gvec{H}\cdot\gvec{\hat x}_{t}^i$ \\
        \hspace{1.3cm}Estimate $\brackets{\hat\vgamma^{i}_{t}}^2$ via~\eqref{eqn:gammaAM} 
        \State Compute $\gvec{\Gamma}_{t}^i=\textnormal{diag}\brackets{\gvec{r}^2+\brackets{\hat\vgamma^i_{t}}^2}$ 
        \State Compute $ \gvec{\hat y}_{t\given{t-1}}^i,\gvec{S}_{t\given{t-1}}^i$ via ~\eqref{eqn:predict2}  with $\gvec{R}=\gvec{\Gamma}_{t}^i$.
        \State \textbf{Update:} Estimate \emph{a posteriori} for $\gvec{\hat x}_{t}^i, \mySigma_{t}^i$ via~\eqref{eqn:update} 
        \EndFor}
    \end{algorithmic}
\end{small}
\end{algorithm}
%
%
\subsection{Discussion}\label{subsec:discussion}
The \ac{nuv} modeling is particularly efficient due to its sparse features, suitable for handling \ac{kf} in presence of outliers, which can be achieved by selecting the prior distribution of $\vgamma^2$. To that end, we opt for a uniform prior for the computations' convenience, which effectively adjusts the overall loss function to accommodate very sparse outliers' detection. Different choices for this prior would lead to alternative loss functions, such as the convex Huber cost function~\cite{Roncetti2009 }. \textcolor{black}{Furthermore, the efficiency of \ac{nuv} modeling comes from being parameter-free and not requiring hyper-parameter tuning, and from a computational perspective, it involves a short iterative process in the \ac{kf} update step, leveraging all observation samples.}

To integrate the \ac{nuv} within the \ac{kf}, we utilize either the \ac{em} or \ac{am} algorithms to estimate the unknown variance of the \ac{nuv}. The estimated result combines to enhance the \ac{kf} update step. The main limitation of the \ac{em} is that it requires the posterior variances of the state $\gvec{x}_t$ in each iteration, which may be infeasible for large problems, while in \ac{am} version it is obtained from the first-order moment. However, the \ac{am} proves more effective compared to \ac{em},  as evident from the empirical evaluation in Section~\ref{sec:NumEval}. 

It is important to note that in certain applications, \ac{em} empirically gives better results than \ac{am}, primarily due to its accounting for the accuracy of the state estimate.
Its simplicity relies on empirical second-order moments, and holds potential for augmentation with trainable data-driven variations of the \ac{kf}, for instance~\cite{Revach2022,ni2022rtsnet}. Such fusion leverages robust filtering in partially known \ac{ss} models and helps manage sensitivity to outliers. 
%
%
\begin{figure*}[t]
\begin{center}
    \begin{subfigure}[h]{0.65\columnwidth}
        \includegraphics[width=0.98\columnwidth]{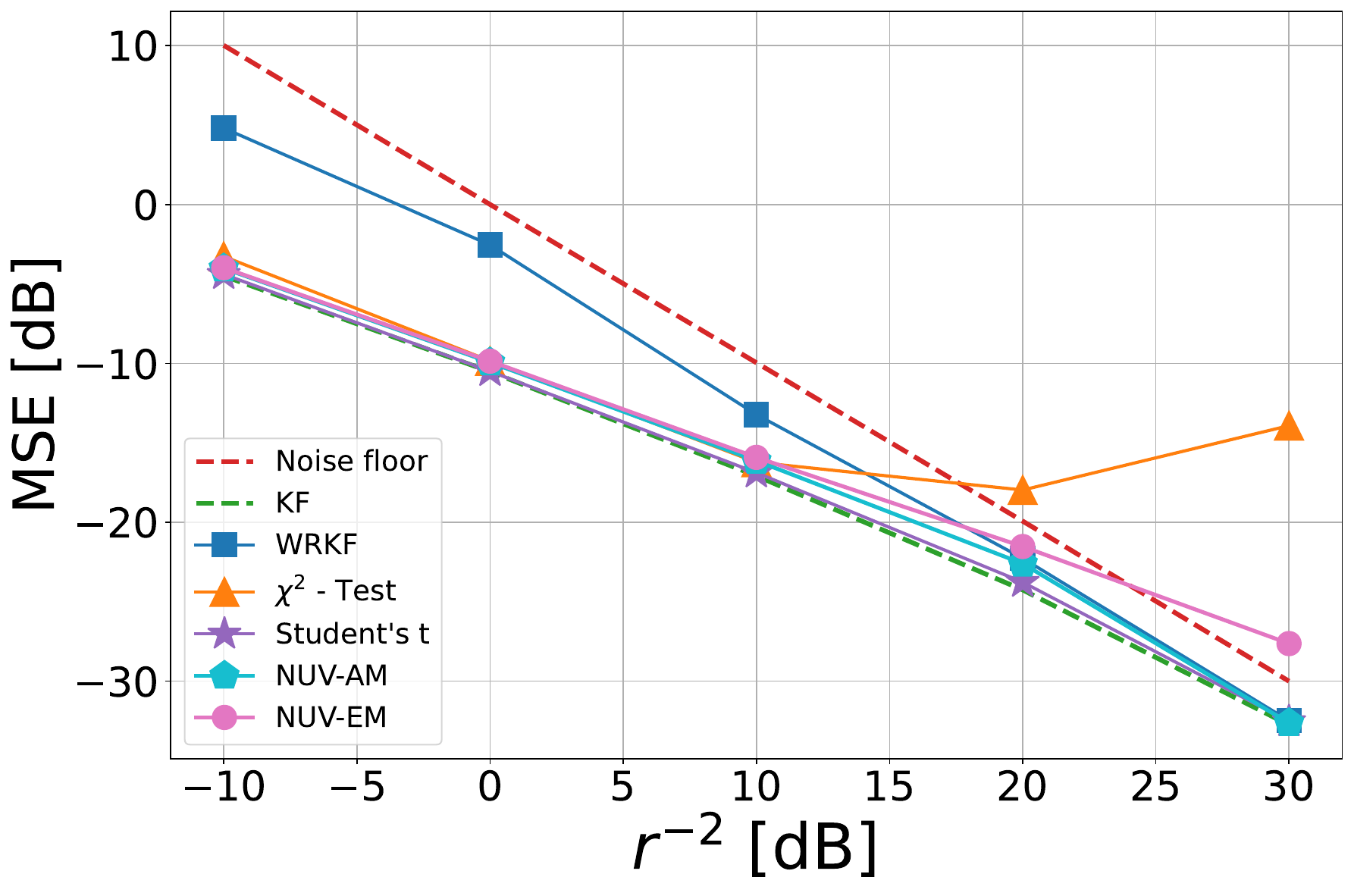}
        \caption{Noisy data clean of outliers}
        \label{fig:MSE noisy}
    \end{subfigure}
    \figSpace
    \begin{subfigure}[h]{0.65\columnwidth}
        \includegraphics[width=0.98\columnwidth]{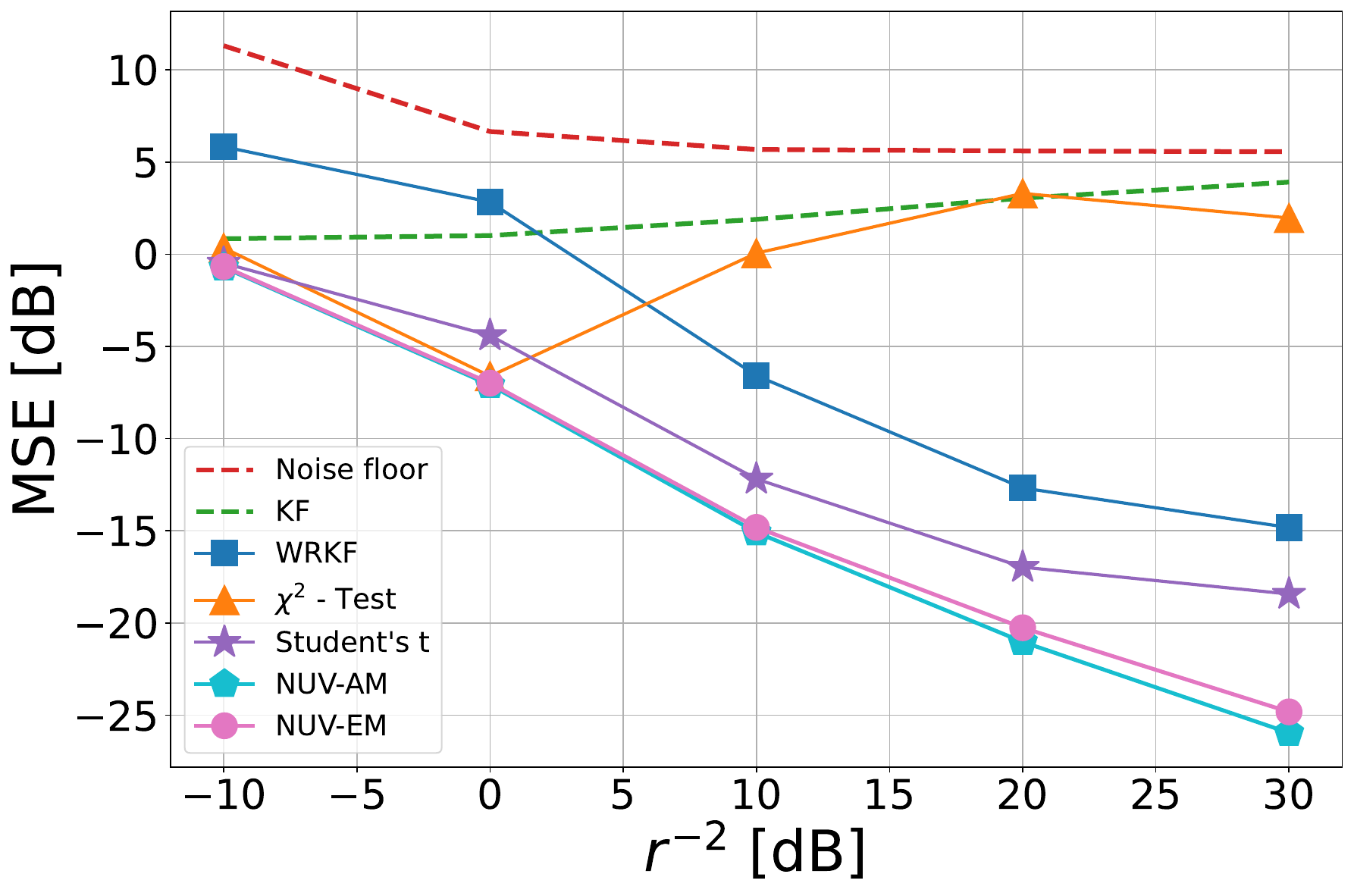}
        \caption{Noisy data with outliers $p=0.2$ $scl=3$}
        \label{fig:MSE noisy with low outliers}
    \end{subfigure}
    \figSpace
    \begin{subfigure}[h]{0.65\columnwidth}
        \includegraphics[width=0.98\columnwidth]{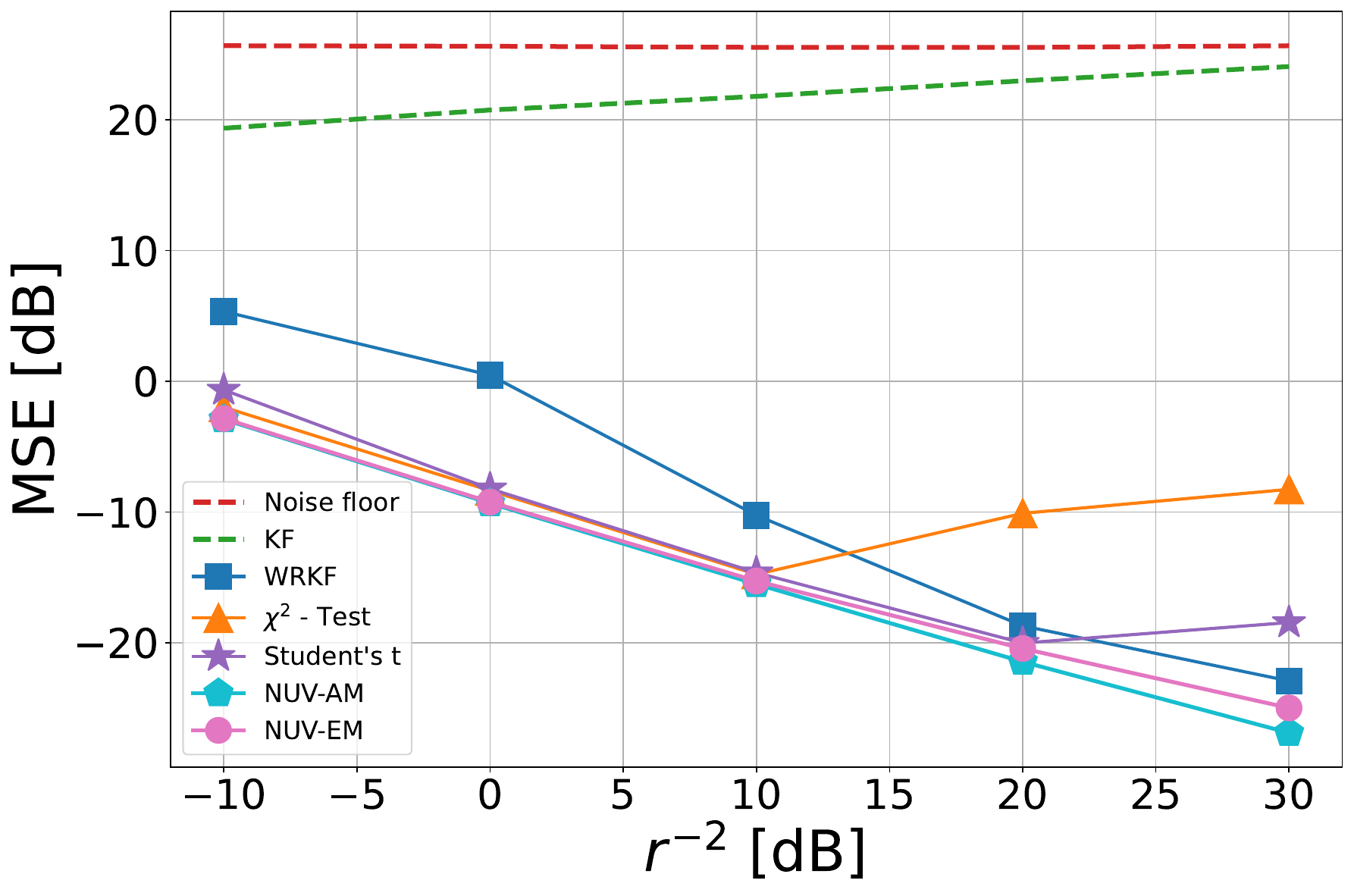}
        \caption{Noisy data with outliers $p=0.2$ $scl=30$}
        \label{fig:MSE noisy with high outliers}
    \end{subfigure}
\caption{Sub-Fig.~\ref{fig:MSE noisy}, ~\ref{fig:MSE noisy with low outliers} and ~\ref{fig:MSE noisy with high outliers} present the  \ac{mse} of the estimated position for the tracking application in the \ac{kf} setup. Our \ac{nuv} methods were compared to different well-established robust \ac{kf} algorithms in the literature.}
\label{fig:synthetic_dateset}
\end{center}
\figSpace
\vspace{-2mm}
\end{figure*}
%
\section{Analysis and Results}\label{sec:NumEval}
In this section, we present a comprehensive assessment of the effectiveness of our proposed approaches: \ac{oikf} based \ac{nuv}-\ac{em}, and \ac{oikf} based \ac{nuv}-\ac{am}, for outlier detection within various \ac{kf} setups. Their performance is evaluated across different outlier intensities and tasks while comparing their effectiveness to other established works in the literature:
\begin{enumerate}[label=(\alph*)]
\vspace{0.5mm}
\item \textbf{Simulations}: Our first experimental study considers a standard localization task with generated data. The synthetic dataset is generated using the \ac{wna}~\cite{Bar2004} model, and the observation signal is subject to varying degrees of outlier corruption. Such models are commonly used in several applications such as navigation and target tracking. 
\vspace{0.5mm}
\item \textbf{\ac{nclt} Dataset}: In our second study, we examine localization use case based on \acl{rw} data - the Michigan \ac{nclt}~\cite{Bianco2016} dataset. Here, we compare our methods with different algorithms for tracking \acl{rw} dynamic data of a moving Segway robot using GNSS noisy measurements. 
\vspace{0.5mm}
\item \textbf{\ac{api} Dataset}: {The third and fourth studies involve another localization use case, based on the \acl{rw} data - the \ac{api}~\cite{Shurin2022, shurin2022quadnet} dataset. Here, we demonstrate the performances of our algorithms in tracking a quadrotor and a marine vessel using GNSS noisy measurements.}
\end{enumerate}
%
\subsection{Simulations}
In this study, we utilize a \ac{kf} algorithm where the state vector is represented by,
\begin{equation}
    \gvec{x}=\begin{bmatrix}
        \gscal{p} & \gscal{v}
    \end{bmatrix}^\top
\end{equation}
where $\gscal{p}$ and $\gscal{v}$  denote the position and velocity states, respectively.
For the experiments that involve generated data, we establish the dynamic \acl{wna} model, followed by a linear \ac{ss} model. For the filtering process, we assume both position and velocity measurements are available, thus the observation matrix is the identity matrix and the observation noise covariance matrix is diagonal; i.e.,
\begin{equation}
    \gvec{H}=
        \begin{pmatrix}
            1 & 0 \\
            0 & 1 
        \end{pmatrix},\quad 
            \gvec{R}=   \gscal{r^2}
        \begin{pmatrix}
            1 & 0 \\
            0 & 1 
        \end{pmatrix}
\end{equation}
The process noise covariance matrix is :
\begin{equation}
    \gvec{Q}=   \gscal{q^2}
    \begin{pmatrix}
        1 & 0 \\
        0 & 1 
    \end{pmatrix}
\end{equation}
where $\gscal{q^2}$ is the process noise variance, set to a constant value of $-10\dB$.

To evaluate our approach, we consider three scenarios for generating measurements: noisy data without outliers, noisy data with mild outliers, and noisy data with significant outliers.
The presence of outliers within the measurement vector is modeled with intensities drawn from a Rayleigh distribution with parameter $\sigma{_u^2}$, where we employ two scale parameters $\sigma{_u^2}=\sbrackets{3,30}$, representing low and high outlier intensities, respectively.
The occurrence of outliers in the dataset is determined using a Bernoulli distribution, $\mathcal{B}(p)$, where we set the probability of an outlier to be $p=0.2$, indicating that roughly $20\%$ of the data points are considered outliers.
In \figref{fig:synthetic_dateset} we compare our proposed \ac{oikf} based on \ac{nuv}-\ac{em} and based on \ac{nuv}-\ac{am} with the following algorithms: classical \ac{kf}, reweighted algorithm (WRKF\cite{Ting2007}), $\chi^2$-test~\cite{Lekkas2015, VanWyk2020} \textcolor{black}{and Student's t-distribution \cite{Agamennoni2012}}. \textcolor{black}{The results are presented in dB units to accentuate the distinctions between the obtained results. The unit conversion to dB is as follows:
\begin{equation}
\textnormal{MSE}\dB = 10\cdot \log_{10}(\textnormal{MSE})
\end{equation}
This decision was motivated by the fact that, in some cases, the differences can be subtle, making them challenging to notice without the emphasis provided by dB units.}

\figref{fig:MSE noisy} presents the results of evaluating the performance of the \ac{nuv}-\ac{am} algorithm on synthetic data that is clean of outliers, wherein the algorithm achieves the optimal minimal \ac{mse} bound by estimating a significant proportion of $\vgamma^2_t$ values as zero. Consequently, the model reverts to the \ac{kf}, \textcolor{black}{which is optimal for data without outliers.} \textcolor{black} 
{However, the student's t outperforms our method because we optimize its hyperparameter such that it tends toward a normally distributed model (as the \ac{kf}), which is also its drawback because it isn't an adaptive algorithm.} In contrast, the $\chi^2$-test is deviating significantly from the performance of the \ac{kf}. This is due to the necessity of determining its confidence level. In cases of low-noise observations, it may incorrectly identify normal samples as outliers, leading to their rejection and rendering the model reliant solely on the prediction step. {The relative efficiencies of our \ac{oikf} and the \ac{kf} can be assessed based on the resulting \ac{mse}. Notably, as depicted in \figref{fig:MSE noisy}, the average ratio of \ac{mse} values between the \ac{kf} and \ac{oikf}-based \ac{nuv}-\ac{em} yields a relative efficiency of $92\%$, while \ac{oikf}-based \ac{nuv}-\ac{am} results in a relative efficiency of $96\%$. This outcome underscores the effectiveness of our algorithms even in the absence of outliers,  due to the fact that a Gaussian framework is maintained.}

When utilizing synthetic data with outliers, as shown in~\figref{fig:MSE noisy with low outliers} and~\figref{fig:MSE noisy with high outliers} with outlier intensities of $3$ and $30$, respectively, the \ac{nuv}-\ac{am} algorithm exhibits superior performance in terms of \ac{mse} as compared to other algorithms, across varying values of observation noise variance $r^2$, and for both high and low outlier intensities. Notably, it demonstrates similar performance to the \ac{nuv}-\ac{em} algorithm and even surpasses it for low observation noise without utilizing a second-order moment as in \ac{em}.  

\begin{figure}[h]
\begin{center}
    \begin{subfigure}[h]{0.49\columnwidth}
        \includegraphics[width=1\columnwidth]{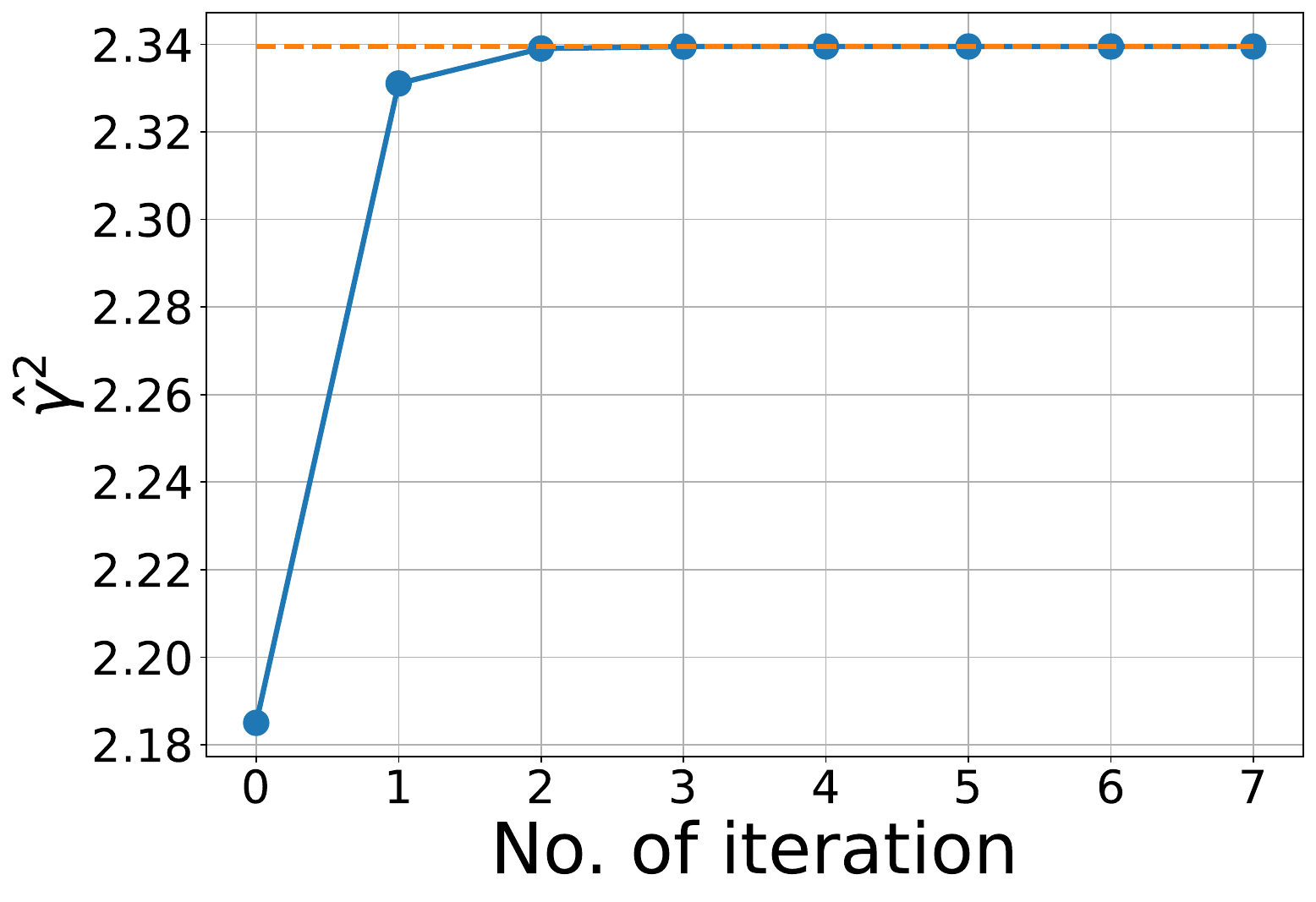}
        \caption{Low outlier intensity}
    \end{subfigure}
    \begin{subfigure}[pt]{0.49\columnwidth}
        \includegraphics[width=1\columnwidth]{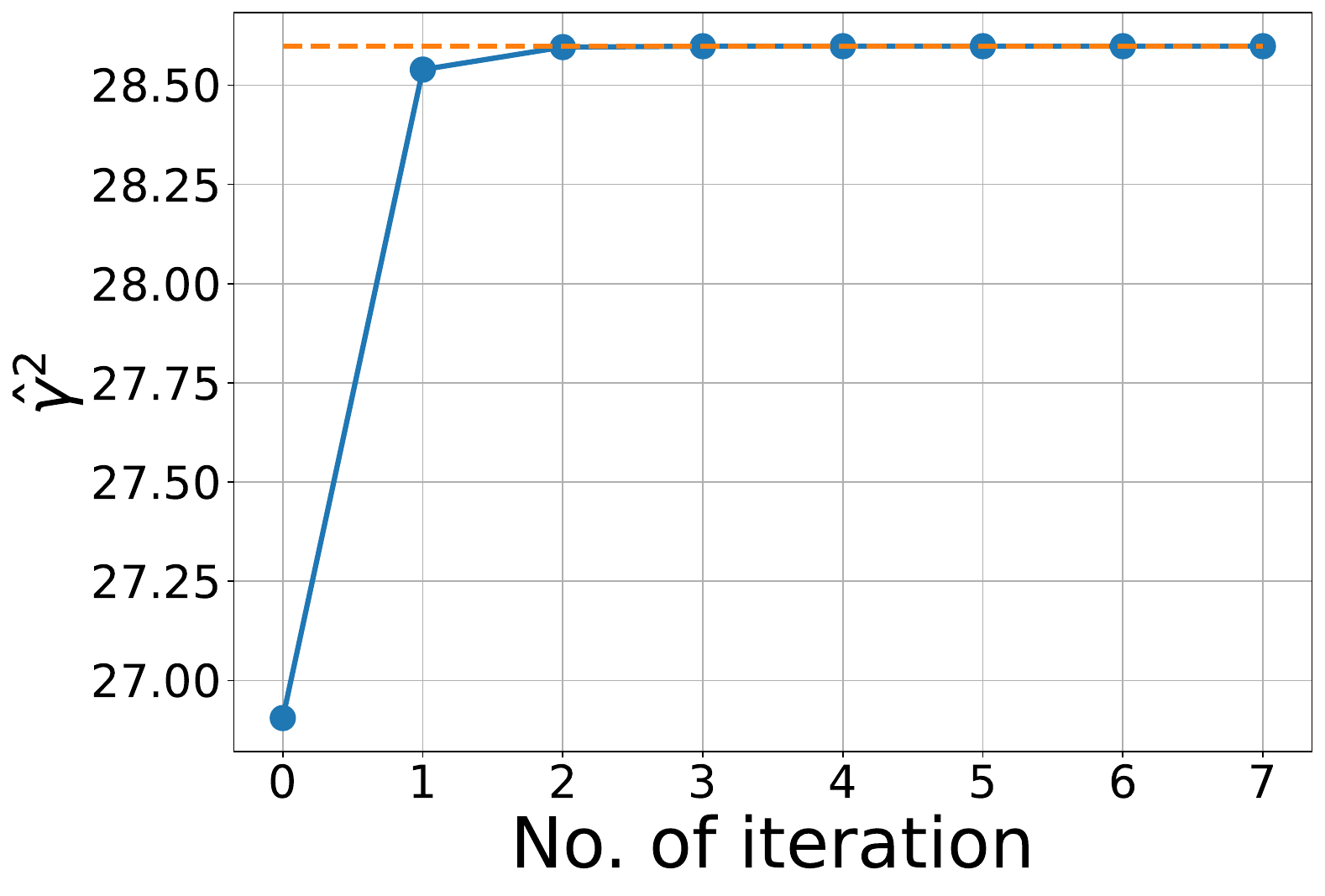}
        \caption{High outlier intensity}
    \end{subfigure}
\caption{Convergence plots of the estimated outliers' variance computed using the NUV-\ac{am} algorithm.}
\label{fig:convergence}
\end{center}
\end{figure}

In~\figref{fig:convergence}, we exhibit the convergence plots of the estimated variance $\hat{\gamma}^2$, using the NUV-\ac{am} algorithm when an outlier was identified.  It is evident that the NUV-AM algorithm achieves rapid convergence after approximately three iterations, regardless of whether the outlier intensities are high or low.

%
%

%
\subsection{NCLT Dataset}\label{subsec:NCLT}
For \acl{rw} data, we make use of the \ac{nclt} dataset~\cite{Bianco2016}. The \ac{nclt} dataset is collected from a session with the date \textit{2013-04-05}, which \ac{gnss} readings sampled at 5[Hz] with a degree of noise and the corresponding ground location information of a Segway robot in motion.
\textcolor{black}{In the simulation setting, we process the measured vehicle position directions independently. To filter out these processes, we employ the dynamic \acl{wna} model~\cite{Bar2004} for each direction separately.}
Since only the \ac{gnss} position is observable in this dataset, the measurement matrix is:
\begin{equation}
    \gvec{H}=
    \begin{bmatrix}
        1 & 0  
    \end{bmatrix},
\end{equation}
and the process and measurement noise covariances are
\begin{equation}
    \gvec{R}=   \gscal{r^2},  \quad 
    \gvec{Q}=   \gscal{q^2}
    \begin{pmatrix}
        1 & 0 \\
        0 & 1 
    \end{pmatrix}
\end{equation}
%

\begin{figure}[H]
\begin{center}
    \begin{subfigure}[pt]{0.49\columnwidth}
        \includegraphics[width=1\columnwidth]{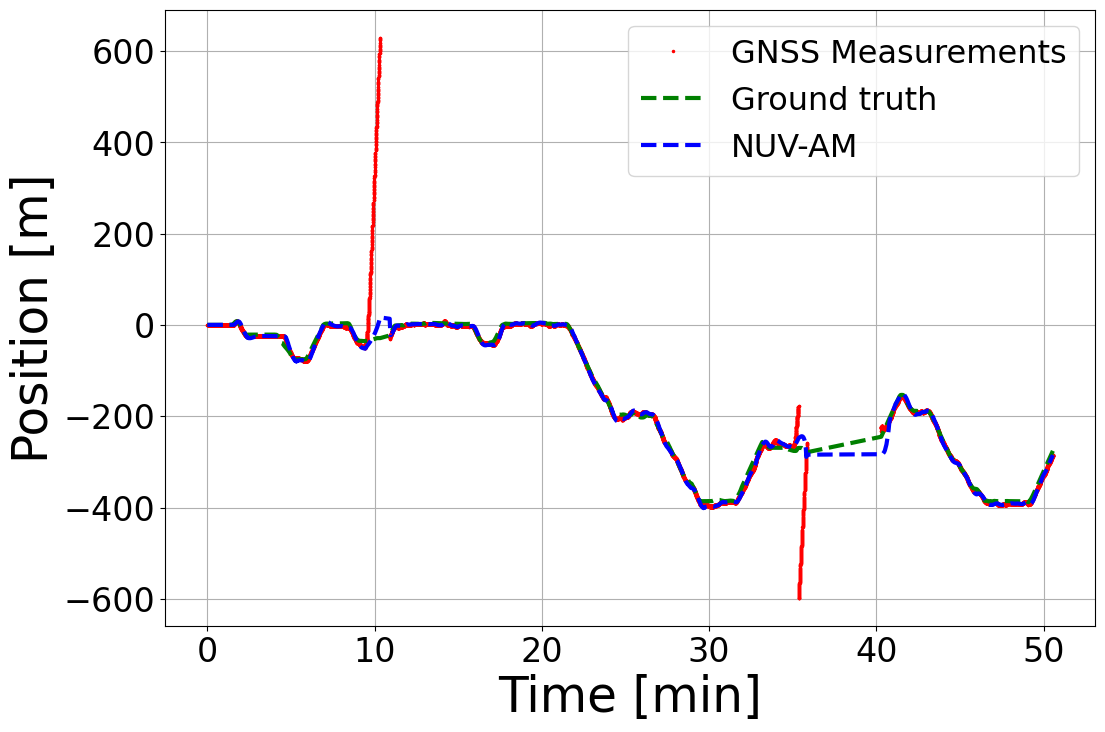}
        \caption{Time interval 0-50[min]}
    \end{subfigure}
    \begin{subfigure}[pt]{0.49\columnwidth}
        \includegraphics[width=1\columnwidth]{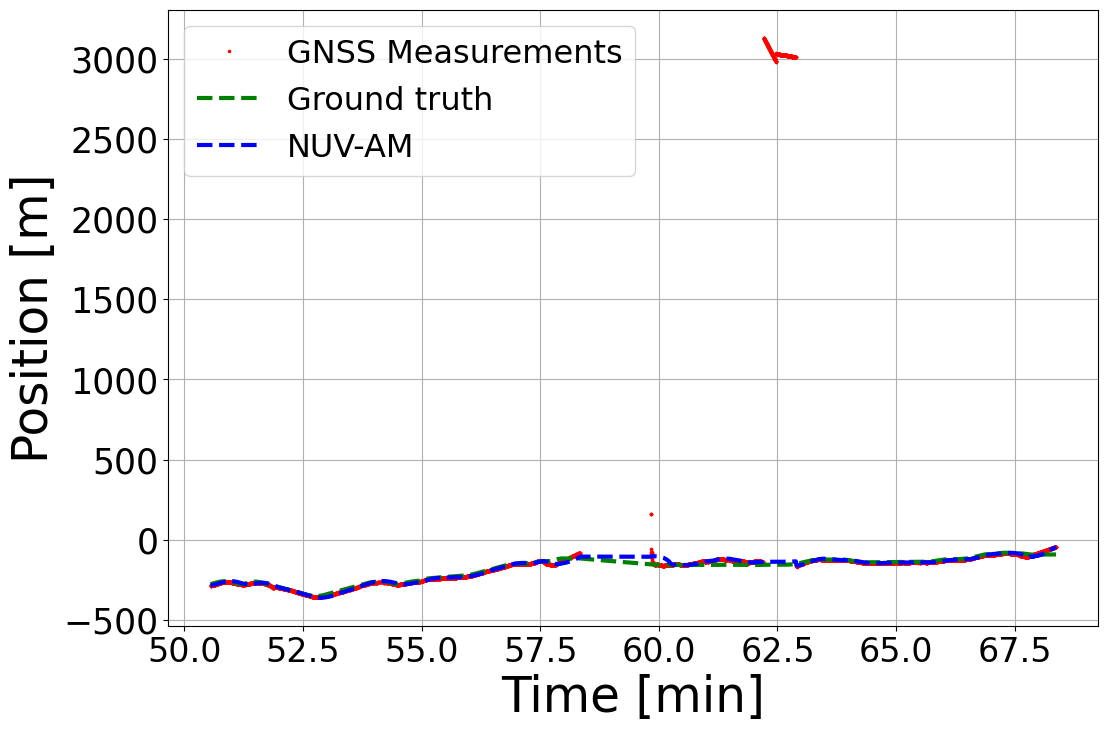}
        \caption{Time interval 50-68[min]}
    \end{subfigure}
\vspace{-1mm}
\caption{The measured vehicle position in \textit{east direction} obtained from the noisy \ac{gnss} \ac{nclt} dataset (red points), is compared to the estimated trajectory by our \ac{nuv}-\ac{am} (blue dashed line), which succeeded in passing the outliers and achieved performance comparable to the ground-truth (green dashed line).}
\label{fig:NCLT_east_dir}
\end{center}
\vspace{-4mm}
\end{figure}

\begin{figure}[H]
\begin{center}
    \begin{subfigure}[pt]{0.49\columnwidth}
        \includegraphics[width=1\columnwidth]{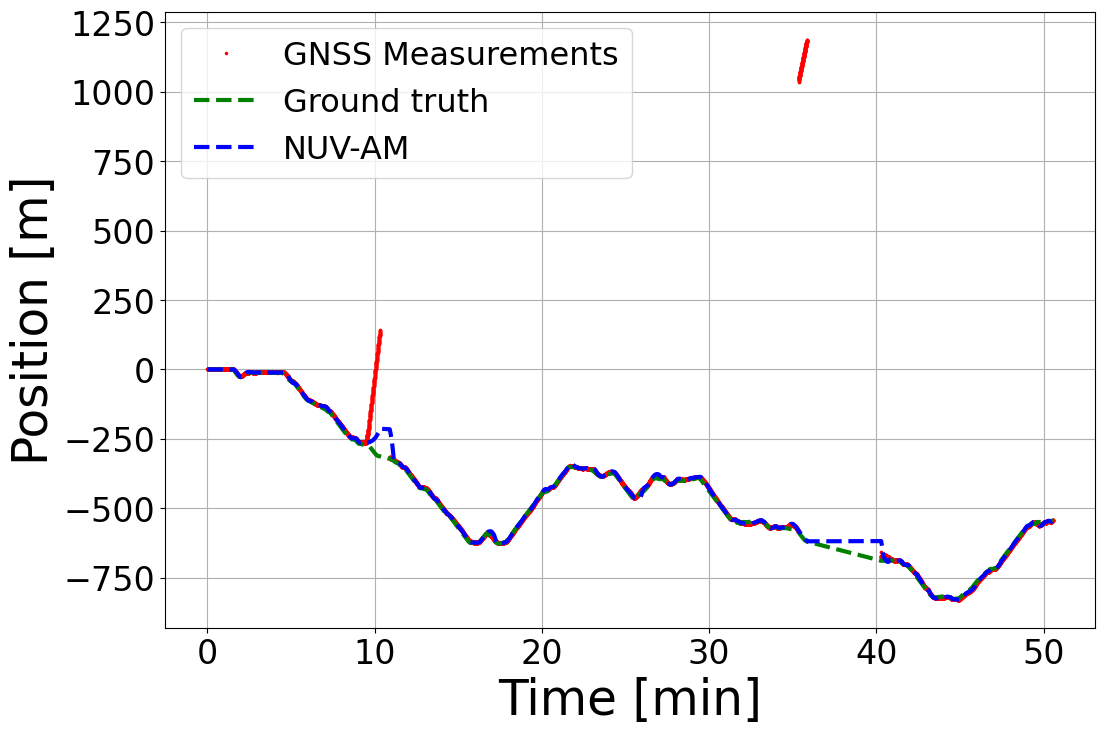}
        \caption{Time interval 0-50[min]}
    \end{subfigure}
    \begin{subfigure}[pt]{0.49\columnwidth}
        \includegraphics[width=1\columnwidth]{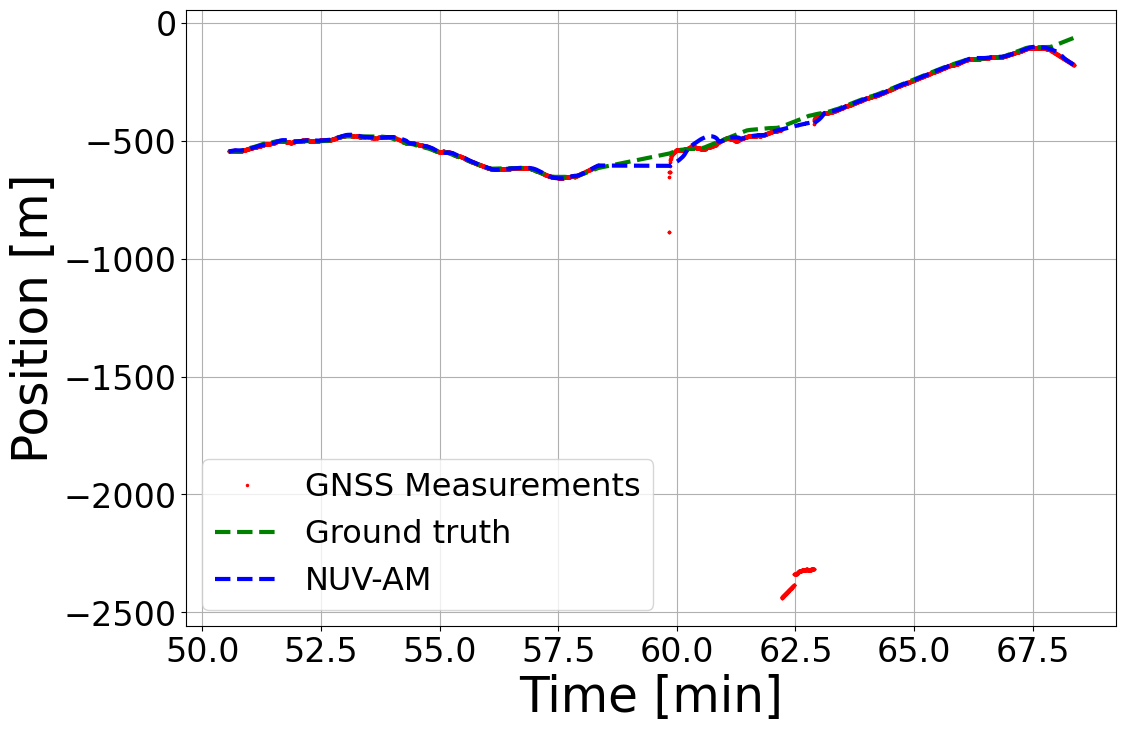}
        \caption{Time interval 50-68[min]}
    \end{subfigure}
\vspace{1mm}
\caption{The measured vehicle position in \textit{north direction} obtained from the noisy \ac{gnss} \ac{nclt} dataset (red points), is compared to the estimated trajectory by our \ac{nuv}-\ac{am} (blue dashed line), which succeeded in passing the outliers and achieved performance comparable to the ground-truth (green dashed line).}
\label{fig:NCLT_north_dir}
\end{center}
\vspace{-1mm}
\end{figure}

%
~\figref{fig:NCLT_east_dir} and~\figref{fig:NCLT_north_dir} depict the trajectory of the Segway in the east and north directions, respectively, when the \ac{gnss} measurements are subjected to outliers, that result in readings deviating significantly from the \ac{gt}. We have divided each trajectory into two time intervals, with the first interval displaying outliers with lower intensity and the second interval with higher intensity.

\begin{figure}[h]
\begin{center}
    \begin{subfigure}{0.49\columnwidth}
        \includegraphics[width=1\columnwidth]{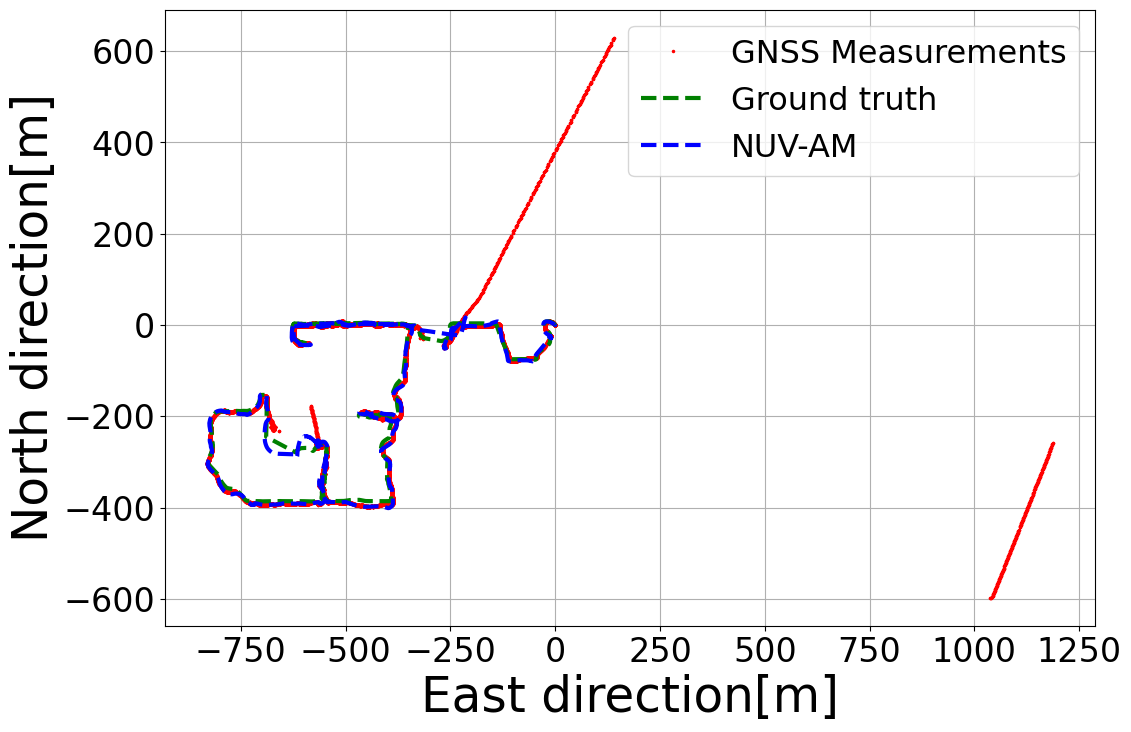}
        \caption{Time interval 0-50[min]}
    \end{subfigure}
    \begin{subfigure}{0.49\columnwidth}
        \includegraphics[width=1\columnwidth]{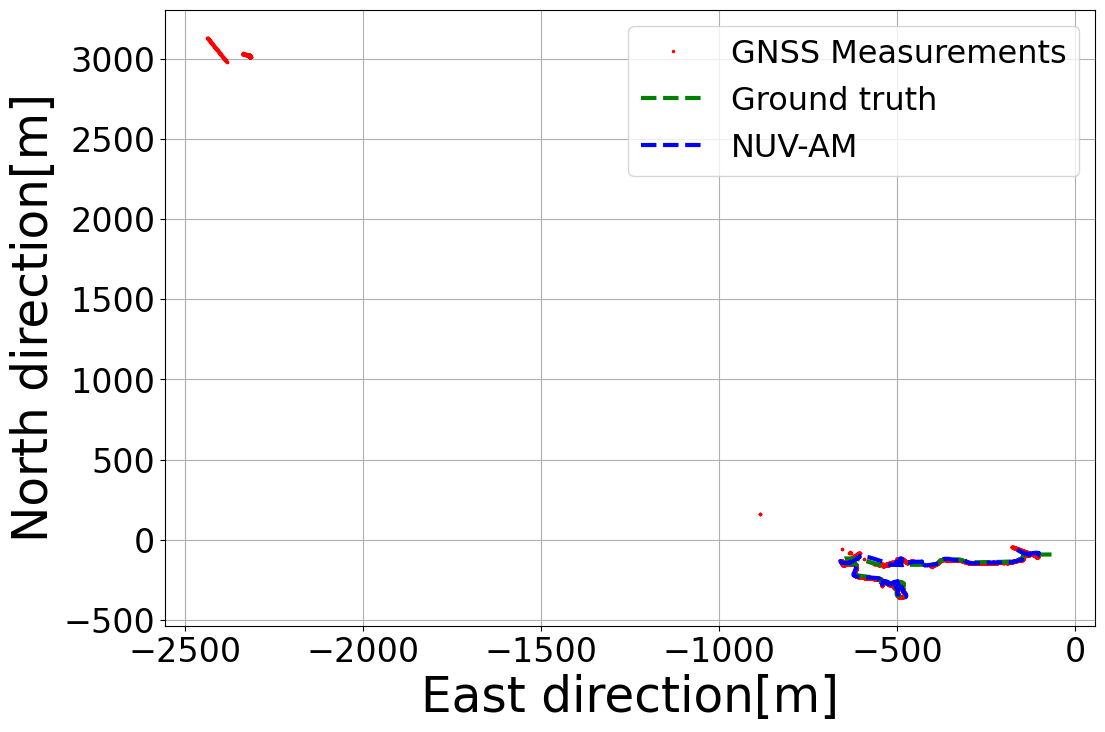}
        \caption{Time interval 50-68[min]}
    \end{subfigure}
\vspace{-2mm}
\caption{The measured vehicle position structure obtained from the noisy \ac{gnss} \ac{nclt} dataset (red points), is compared to the estimated trajectory by our \ac{nuv}-\ac{am} (blue dashed line), which succeeded in passing the outliers and achieved performance comparable to the ground-truth (green dashed line).}
\label{fig:NCLT_eastVSnorth}
\end{center}
\end{figure}

In addition, \figref{fig:NCLT_eastVSnorth} presents the spatial trajectory of the moving Segway in both directions. In the $x$-axis, we depict the trajectory from~\figref{fig:NCLT_east_dir}, while  the $y$-axis depicts the trajectory from~\figref{fig:NCLT_north_dir}.
%
\vspace{2mm}
Our analysis focuses on evaluating the effectiveness of \ac{nuv}-\ac{am} in estimating position from real-world data and removing outliers reliably. \textcolor{black}{In Table~\ref{tbl:NCLT results} we compare our proposed \ac{oikf} based on \ac{nuv}-\ac{em} and based on \ac{nuv}-\ac{am} with the following algorithms: classical \ac{kf}, reweighted algorithm (ORKF\cite{Agamennoni2011}), $\chi^2$-test~\cite{Lekkas2015, VanWyk2020} and Student's t-distribution \cite{Agamennoni2012}}. Additionally, Table~\ref{tbl:NCLT results} presents the \ac{rmse} and \ac{mse} for each algorithm, while the process noise $\gscal{q}^2$ and observation noise variance $\gscal{r}^2$ of each algorithm are optimized separately through grid search to yield the lowest \ac{mse}.
As shown in Table~\ref{tbl:NCLT results}, \ac{oikf} with both \ac{nuv}-\ac{em} and \ac{nuv}-\ac{am} has the lowest estimation errors in both directions, with \ac{nuv}-\ac{am} performing slightly better when it coincides with \ac{nuv}-\ac{em}, even without utilizing the second-order moment.

\vspace{2mm}
{In terms of computation time, algorithms that combine \ac{kf} with outlier detection techniques achieve higher accuracy but at the cost of longer computation runtimes, as outlier detection is applied. Compared to other algorithms, the $\chi^2$-test stands out as the shortest, stemming from the fact that it only detects and rejects outliers during the prediction step. On the other hand, a more sophisticated technique, such as the ORKF, requires more time due to its increased computational complexity, which requires tuning multiple parameters. \textcolor{black}{In contrast, the runtime computation for the student's t distribution is significantly shorter compared to the ORKF because it requires tuning only one parameter.} Our algorithms present relatively shorter computation times among outlier detection and weighting algorithms, coupled with low MSE, making them suitable for real-time tasks. Additionally, in comparison to our other suggested method, \ac{nuv}-based \ac{em}, \ac{nuv}-based \ac{am} showcases an almost $40\%$ reduced runtime.}
\vspace{5mm}
%
%
\input{Tab_NCLT_V2}
%

%
\subsection{API Dataset}
\vspace{3mm}
To emphasize the versatility and robustness of our approaches in tracking \acl{rw} dynamic data of different platforms, which may be corrupted by various types of outliers, we evaluate the \ac{api} dataset~\cite{Shurin2022}. The \ac{api} dataset is collected from the \textit{MATRICE 300 quadrotor} platform containing \ac{gnss} RTK reading sampled at $10[Hz]$ {and from a marine vessel named \textit{"Shikmona"} containing motion reference units (MRU) with \ac{gnss} RTK receiver, sampled at $100[Hz]$}. 

\newpage
Their trajectories were populated with generated outliers, sampled with intensity from a Rayleigh distribution, while their time steps within the data were drawn from a Bernoulli distribution. To filter out this process, we use the same model and parameters as in \ssecref{subsec:NCLT}.

\begin{figure}[H]
\begin{center}
    \begin{subfigure}{0.49\columnwidth}
        \includegraphics[width=1\columnwidth]{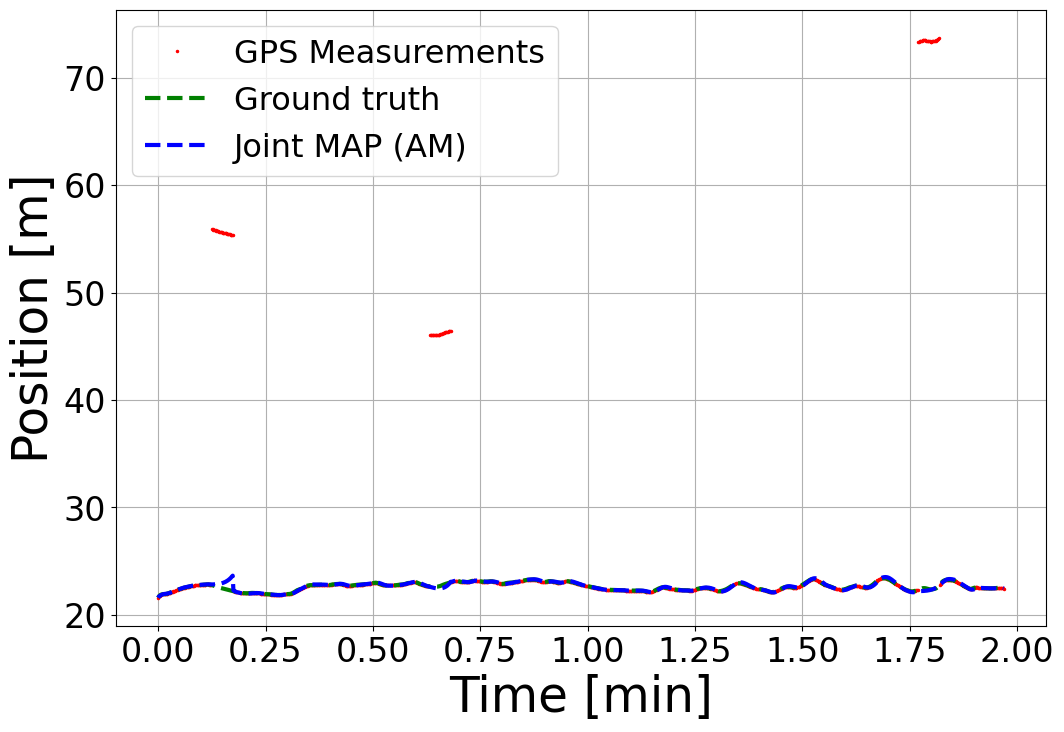}
        \caption{Horizontal direction}
    \end{subfigure}
    \begin{subfigure}{0.49\columnwidth}
        \includegraphics[width=1\columnwidth]{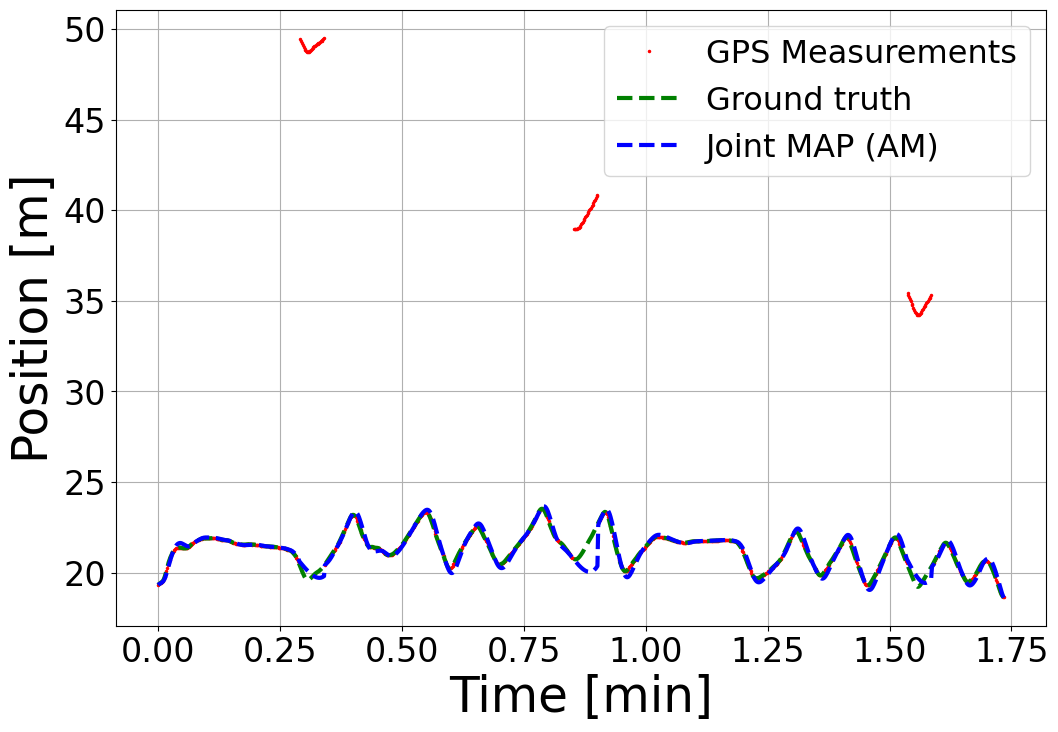}
        \caption{Vertical direction} 
    \end{subfigure}
\vspace{1mm}
\caption{The measured quadrotor trajectory
obtained from the noisy \ac{gnss} measurements (red points),
with the estimated trajectory by our NUV-AM (blue dashed
line), and ground-truth (green dashed line).}
\label{fig:API}
\end{center}
\end{figure}

It is evident from~\figref{fig:API} that our \ac{oikf} based on \ac{nuv}-\ac{am} algorithm accurately estimates the position, effectively handling all outliers in both \textit{vertical} and \textit{horizontal} directions of the quadrotor trajectory. 
%

\textcolor{black}{In Table~\ref{tbl:API results} we compare our proposed \ac{oikf} based on \ac{nuv}-\ac{em} and based on \ac{nuv}-\ac{am} with the following algorithms: classical \ac{kf}, reweighted algorithm (ORKF\cite{Agamennoni2011}) and $\chi^2$-test~\cite{Lekkas2015, VanWyk2020}}. Additionally, Table~\ref{tbl:API results} presents the \ac{rmse} and \ac{mse} for each algorithm, while the process noise $\gscal{q}^2$ and observation noise variance $\gscal{r}^2$ of each algorithm are optimized at the same procedure as in the Table~\ref{tbl:NCLT results}. As shown in Table~\ref{tbl:API results}, \ac{oikf} with both \ac{nuv}-\ac{em} and \ac{nuv}-\ac{am} has the lowest estimation errors for both directions and emphasize the use in \ac{nuv} method for outlier detection.

\begin{figure}[H]
\begin{center}
        \includegraphics[width=0.52\columnwidth]{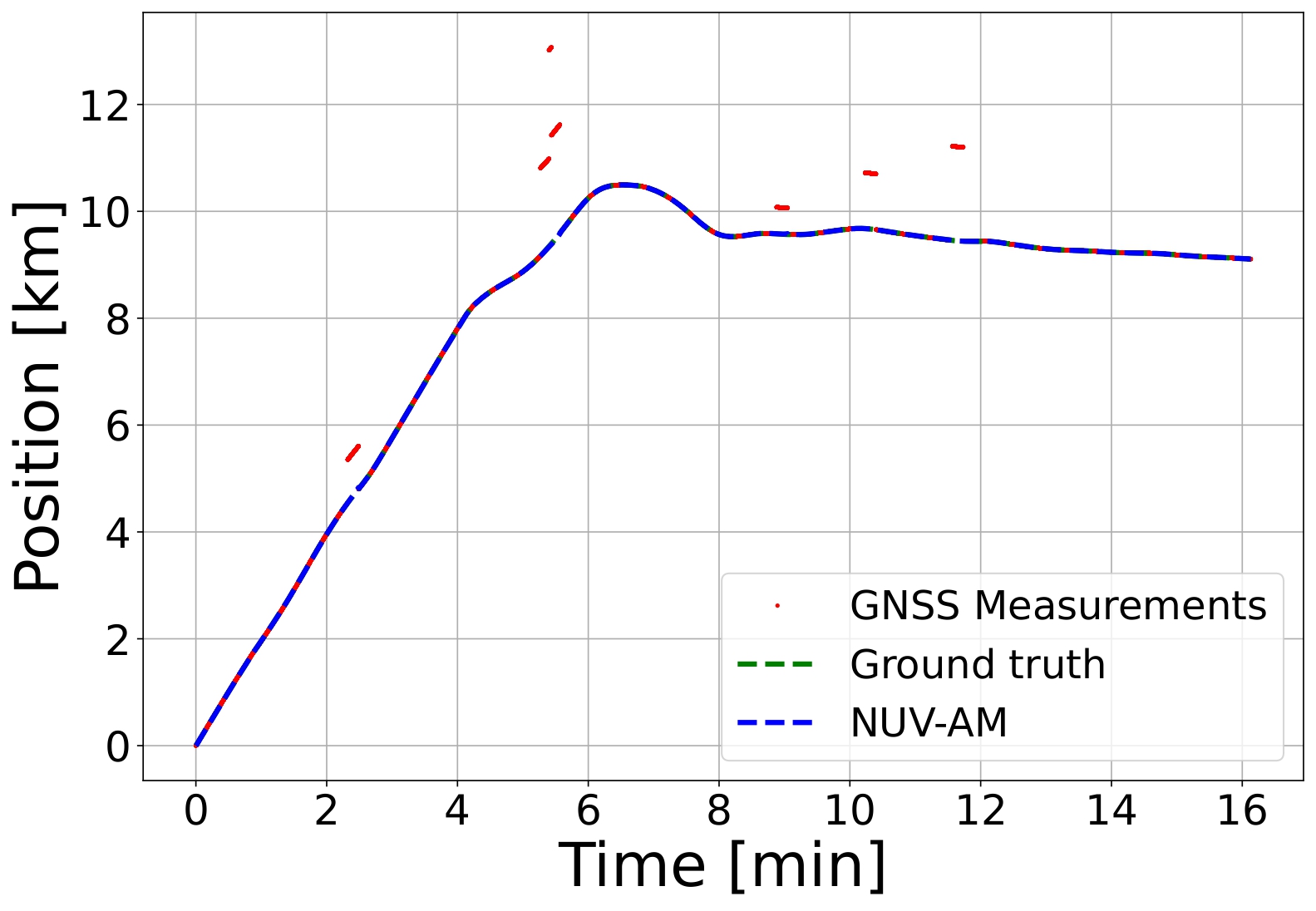} 
\caption{{The measured \textit{"Shikmona"} marine vessel trajectory obtained from the noisy \ac{gnss} measurements (red points), with the estimated trajectory by our \ac{nuv}-\ac{am} (blue dashed line), and ground-truth (green dashed line).}}
\label{fig:MV}
\end{center}
\end{figure}

%
%
\input{API_Tab}
{\figref{fig:MV} demonstrates the performance of our \ac{oikf}-based \ac{nuv}-\ac{am} in estimating the position of the \textit{"Shikmona"} marine vessel. The examined trajectory includes straight line segments and turns. As can be seen, \ac{oikf}-based \ac{nuv}-\ac{am} successfully tracks the ground truth and surpasses outliers, even during turns.} \textcolor{black}{Table~\ref{tbl:MV} underscores the superior performance of our proposed \ac{oikf} based on \ac{nuv}-\ac{em} and based on \ac{nuv}-\ac{am} compared to the following algorithms: classical \ac{kf}, reweighted algorithm (ORKF\cite{Agamennoni2011}) and $\chi^2$-test~\cite{Lekkas2015, VanWyk2020}}, revealing significantly low \ac{rmse} and \ac{mse} values for both our \ac{nuv}-\ac{em} and \ac{nuv}-\ac{am} algorithms.
\vspace{1mm}
%
%
%
\input{API_MV_Tab}
%
%
\section{Conclusion}\label{sec:Conclusions}
%
In this work, we have proposed an innovative outlier-insensitive \ac{kf} that offers improved performance to tackle the problem of state estimation in the presence of outliers. Based on Bayesian learning concepts, we model the outlier as \ac{nuv} and estimate the unknown variance using either \ac{em} or \ac{am} algorithms, resulting in sparse outlier detection. Both algorithms are parameter-free and amount essentially to a short iterative process during the update step of the \ac{kf}.
Our numerical study demonstrates the effectiveness of our algorithms and highlights the robustness and wide applicability in addressing a variety of applications.  We demonstrate superior performances competing with other algorithms in terms of \ac{mse} and \ac{rmse} across synthetic and \acl{rw} datasets. These findings emphasize the robustness and accuracy of our \ac{oikf} approach, making it especially suitable for systems reliant on high-quality sensory data.
%


\end{document}

%% file: Tab_NCLT_V2.tex
\begin{table}[H]
\scriptsize{
\captionsetup{justification=centering,margin=1cm}
\caption{Position error for optimal values of $\gscal{r}^2$ and $\gscal{q}^2$ for \ac{nclt} dataset}
\vspace{-2mm}
\begin{center}
{
\setlength\tabcolsep{4.5pt}
\begin{tabular}{|m{1.2cm}|m{0.95cm}|m{0.85cm}|m{0.95cm}|m{0.85cm}|C{0.85cm}|}
\hline
& \multicolumn{2}{c|} { {\cellcolor{gray} 
North direction}} &  \multicolumn{2}{c|}{{\cellcolor{gray} East direction }}& {\cellcolor{lightgray}Runtime}\\
&{\cellcolor{lightgray}RMSE[m]} & { {\cellcolor{lightgray}MSE$\dB$}} &  {\cellcolor{lightgray}RMSE[m]} & {{\cellcolor{lightgray}MSE$\dB$}}  & {\cellcolor{lightgray}[ms]} \\ 
\hline
{Noisy GNSS} &  349.3   &  50.8  & 266.1  &  48.5  & -    \\
{KF}   &  92.3  &  39.3  & 164.4 & 44.3  & 0.05 \\
{ORKF}      &  27.7   & 28.8   &  28   & 28.9 &  2.8  \\
\acl{ch2t} &    12.3  &    21.8 &    14.2  &   23   & 0.1 \\
{Student's t}    &    11.5  &    21.2 &    13.8  &   22.8   & 0.5 \\
{NUV-AM}    &  \textbf{10.4} &     \textbf{20.3}  &    \textbf{13}      &   \textbf{22.3}    & \textbf{0.3}\\
{NUV-EM}  &   \textbf{10.3}       &     \textbf{20.3} &    \textbf{13} &   \textbf{22.3} & 0.4  \\
\hline      
\end{tabular}  
\vspace{-1mm}
\label{tbl:NCLT results}
}
\end{center}}
\end{table}

%% file: API_Tab.tex
\begin{table}[h]
\scriptsize{
\captionsetup{justification=centering,margin=0.9cm}
\caption{Position error for optimal values of $\gscal{r}^2$ and $\gscal{q}^2$ for \ac{api} dataset {- quadrotor recordings}}
\vspace{-2mm}
\begin{center}
{
\setlength\tabcolsep{4.5pt}
\begin{tabular}{|m{1.6cm}|m{0.95cm}|m{0.95cm}|m{0.95cm}|m{0.95cm}|}
\hline
& \multicolumn{2}{c|} { {\cellcolor{gray} 
Horizontal direction}} &  \multicolumn{2}{c|}{{\cellcolor{gray} Vertical direction }}\\
&{{\cellcolor{lightgray}RMSE[m]}} & { {\cellcolor{lightgray}MSE$\dB$}} &  { {\cellcolor{lightgray}RMSE[m]}} & { {\cellcolor{lightgray}MSE$\dB$}}  \\
\hline
{Noisy GNSS} &  10.4   &   20.3  &  6.4   &   16.1   \\
{KF}   & 0.9   &    -1.2   & 1.8    &    5.3 \\
{ORKF}      & 0.8    &   -1.9   &  1   &   -0.4  \\
\acl{ch2t} &    0.3   &   -11.4   &     0.5    &   -6.3   \\
{NUV-AM}    &  \textbf{0.1} & \textbf{-17.9}  &    \textbf{0.3}      &   \textbf{-10}   \\
{NUV-EM}  &   \textbf{0.1}       &     \textbf{-17.9} &     \textbf{0.3}    &   \textbf{-10.2}  \\
\hline      
\end{tabular}  
\label{tbl:API results}
}
\vspace{-3mm}
\end{center}}
\end{table}

%% file: API_MV_Tab.tex
\begin{table}[h]
\scriptsize{
\captionsetup{justification=centering,margin=0.8cm}
\caption{Position error for optimal values of $\gscal{r}^2$ and $\gscal{q}^2$ for \ac{api} dataset {- marine vessel recordings}}
\vspace{-3mm}
\begin{center}
{
\setlength\tabcolsep{5pt}
\begin{tabular}{|m{1.8cm}|C{1.5cm}|C{1.5cm}|}
\hline
&{{\cellcolor{lightgray}RMSE[m]}} & { {\cellcolor{lightgray}MSE$\dB$}}  \\
\hline
{Noisy GNSS} &  373   &   51    \\
{KF}   & 267   &    48    \\
{ORKF}      &  48    &   34    \\
\acl{ch2t} &    6   &   16     \\
{NUV-AM}    &  \textbf{2.4} & \textbf{7.52}    \\
{NUV-EM}  &   \textbf{2.4}       &     \textbf{7.52} 
\\
\hline      
\end{tabular}  
\label{tbl:MV}
}
\vspace{-4mm}
\end{center}}
\end{table}